\documentclass[cernpreprint,USenglish]{na61doc}
\usepackage{amsmath}
\usepackage{url}
\usepackage{multirow}
\usepackage{makecell}

\usepackage{colortbl}
\definecolor{darkred}{rgb}{0.5,0,0}
\definecolor{darkblue}{rgb}{0,0,0.5}
\definecolor{firebrick}{rgb}{0.75,0.125,0.125}
\definecolor{darkgreen}{rgb}{0,0.5,0}

\usepackage[colorlinks=true,linkcolor=firebrick,citecolor=darkgreen,urlcolor=darkblue]{hyperref}

\usepackage{xspace}
\usepackage{enumerate}

\ShineTitle{Measurements of multiplicity fluctuations of identified hadrons
         in inelastic proton-proton interactions at the CERN Super Proton Synchrotron}

\PreprintIdNumber{CERN-EP-2020-142}

\ShineJournal{}

\ShineJournalRef{}
\ShineDOI{}

\ShineAbstract{
Measurements of multiplicity fluctuations of
identified hadrons produced in inelastic p+p interactions at 31, 40, 80,
and 158~\GeVc beam momentum are presented. Three different
measures of multiplicity fluctuations are used:
the scaled variance $\omega$ and strongly intensive measures $\Sigma$ and
$\Delta$.
These fluctuation measures involve second and first moments of joint multiplicity distributions.
Data analysis is preformed using
the Identity method which corrects for incomplete particle identification.
Strongly intensive quantities are calculated in order to allow for a direct comparison to
corresponding results on nucleus-nucleus collisions.
The results for different hadron types are shown as a function of collision energy.
A comparison with predictions of  string-resonance Monte-Carlo models: \Epos, \Smash and \Venus, is also presented.
}

\newcommand{\eV}{\ensuremath{\mbox{e\kern-0.1em V}}\xspace}
\newcommand{\GeV}{\ensuremath{\mbox{Ge\kern-0.1em V}}\xspace}
\newcommand{\MeV}{\ensuremath{\mbox{Me\kern-0.1em V}}\xspace}
\newcommand{\GeVc}{\ensuremath{\mbox{Ge\kern-0.1em V}\!/\!c}\xspace}
\newcommand{\GeVcc}{\ensuremath{\mbox{Ge\kern-0.1em V}\!/\!c^2}\xspace}
\newcommand{\AGeV}{\ensuremath{A\,\mbox{Ge\kern-0.1em V}}\xspace}
\newcommand{\AGeVc}{\ensuremath{A\,\mbox{Ge\kern-0.1em V}\!/\!c}\xspace}
\newcommand{\MeVc}{\ensuremath{\mbox{Me\kern-0.1em V}/c}\xspace}

\newcommand{\cm}{\ensuremath{\mbox{cm}}\xspace}

\newcommand{\dd}{\ensuremath{{\text{d}}}\xspace}
\newcommand{\dedx}{\ensuremath{\dd E\!/\!\dd x}\xspace}

\newcommand{\pt}{\ensuremath{p_{\text{T}}}\xspace}

\newcommand{\pbar}{\ensuremath{\overline{p}}\xspace}
\newcommand{\pim}{\ensuremath{\pi^-}\xspace}
\newcommand{\pip}{\ensuremath{\pi^+}\xspace}
\newcommand{\km}{\ensuremath{K^-}\xspace}

\newcommand{\pp}{\mbox{\textit{p+p}}\xspace}

\newcommand{\coordinate}[1]{{\fontfamily{lmss}\selectfont#1}}

\newcommand{\Smash}{{\scshape Smash}\xspace}
\newcommand{\SmashLong}{{\scshape Smash1.5}\xspace}

\newcommand{\Venus}{{\scshape Venus}\xspace}
\newcommand{\VenusLong}{{\scshape Venus4.12}\xspace}

\newcommand{\Epos}{{\scshape Epos}\xspace}
\newcommand{\EposLong}{{\scshape Epos1.99}\xspace}

\newcommand{\CernVM}{\textsc{Cern\-\kern-0.05emVM}\xspace}

\begin{document}

\maketitle

\section{Introduction}

This paper presents experimental results on event-by-event
fluctuations of multiplicities of identified particles produced in inelastic proton-proton (p+p) interactions
at 31, 40, 80, and 158~\GeVc $(\!\sqrt{s_{\mathrm{NN}}}$ = 7.6, 8.7, 12.3, 17.3~\GeV).
The measurements were performed by the multi-purpose
\mbox{\NASixtyOne}~\cite{Abgrall:2014xwa} experiment at
the CERN Super Proton Synchrotron (SPS) in 2009.
They are part of the strong interactions programme devoted to the study of the
properties of the onset of deconfinement and search for the
critical point of strongly interacting matter. Within this program, 
a two dimensional scan in collision energy and size of colliding nuclei
was performed~\cite{Aduszkiewicz:2642286}.

An interpretation of the experimental results on nucleus-nucleus (A+A)
collisions relies to a large extent on a comparison with the
corresponding data on p+p and p+A interactions.
In addition models of nucleus-nucleus collisions are often tuned based on results on
p+p interactions. However, available results on fluctuations of identified hadrons in these reactions are sparse.
Moreover, fluctuation measurements cannot be corrected in a model independent manner
for partial phase-space acceptance. Thus all measurements of
the scan should be performed in the same phase space region.
This motivated the \mbox{\NASixtyOne} Collaboration to analyse data on p+p interactions with respect to
fluctuations using the same experimental methods, acceptance and measures as used to study 
nucleus-nucleus collisions.  

Fluctuations in A+A collisions are susceptible
to two trivial sources: the finite and fluctuating number
of produced particles and event-by-event fluctuations of
the collision geometry. Suitable statistical tools have
to be chosen to extract the fluctuations of interest. 
In this publication three different event-by-event fluctuation
measures are used: the scaled variance $\omega$, the $\Delta$ and
$\Sigma$ measures introduced in Refs.~\cite{Gorenstein:2011vq, Gazdzicki:2013ana}. All of them were already successfully utilized by the NA49 experiment at the CERN SPS, see e.g.
Refs.~\cite{Anticic:2003fd, Anticic:2008aa, Alt:2004ir, Anticic:2013htn, Alt:2006jr,Alt:2007jq,Anticic:2015fla} and the \NASixtyOne collaboration, see e.g.
Ref.~\cite{Aduszkiewicz:2015jna}.

Experimental measurements of multiplicity distributions of identified
hadrons are challenging because it is very difficult to identify a particle with sufficient
precision. In this paper the \textit{Identity method}~\cite{Gazdzicki:2011xz, Gorenstein:2011hr, Rustamov:2012bx,Pruneau:2017fim,Mackowiak-Pawlowska:2017dma, Pruneau:2018glv, Arslandok:2018pcu} is employed to circumvent this problem. The Identity method has already been successfully used in the past by 
collaborations NA49~\cite{Anticic:2013htn}, \mbox{\NASixtyOne}~\cite{Mackowiak-Pawlowska:2013caa}, and ALICE~\cite{Acharya:2017cpf, Rustamov:2017lio,Acharya:2019izy}.

The paper is organized as follows. 
In Sec.~\ref{sec:strong_meas} intensive and strongly intensive
measures of fluctuations used in this analysis are introduced
and briefly discussed. 
The Identity method which allows to take into account the  incomplete particle identification is presented in Sec.~\ref{sec:identity} and Appendix~\ref{sec:app2}.
The \NASixtyOne set-up and the data reconstruction method are presented in Secs.~\ref{sec:setup} 
and~\ref{sec:reconstruction}, respectively. 
The data analysis procedure is introduced in Secs.
~\ref{sec:cuts} and~\ref{sec:identity_analysis}. Applied corrections and remaining uncertainties are presented in Sec.~\ref{sec:other}.
Results on the collision energy dependence of multiplicity fluctuations of identified hadrons
in inelastic \pp collisions at 31, 40, 80, and 158~\GeVc beam momentum are presented, discussed and compared with model predictions
in Sec.~\ref{sec:results}. 

Throughout this paper the rapidity is calculated in the
collision center of mass system: $y$ = atanh($\beta_L$), where
$\beta_L = p_L/E$ is the longitudinal (\coordinate{z}) component of the velocity, $p_L$
and $E$ are particle longitudinal momentum and energy given in
the collision center of mass system. The transverse component of the momentum is denoted as $p_T$ and 
the azimuthal angle $\phi$ is the
angle between the transverse momentum vector and the horizontal
(\coordinate{x}) axis. Total momentum in the laboratory system is denoted as $p_{\mathrm{lab}}$ and electric charge is denoted as $q$. The collision energy per
nucleon pair in the center of mass system is denoted as $\sqrt{s_{\emph{NN}}}$.

	\section{Intensive and strongly intensive measures of multiplicity
	and particle type fluctuations}\label{sec:strong_meas}

\subsection{Intensive quantities}

Measures of multiplicities and fluctuations are called intensive when they are \textit{independent} of
the volume ($V$) of systems modelled by the ideal Boltzmann grand canonical ensemble (IB-GCE).
In contrast, extensive quantities
(for example mean multiplicity or variance of the multiplicity
distribution) are proportional to the system volume within IB-GCE.
One can also extend the notion of intensive and extensive quantities to
the Wounded Nucleon Model (WNM)~\cite{Bialas:1976ed}, where the intensive
quantities are those which are independent
of the number of wounded nucleons ($W$), and extensive those
which are proportional to the number of wounded nucleons. Here it is assumed that the number of wounded nucleons is the same for all collisions. 
The ratio of two extensive quantities is an intensive quantity~\cite{Gorenstein:2011vq}.
Therefore, the ratio of mean multiplicities $N_{a}$ and $N_{b}$,
as well as the scaled variance of the multiplicity distribution
$\omega[a] \equiv (\langle N_a^2 \rangle - \langle N_a \rangle ^2)/
\langle N_a \rangle$, are intensive measures. As a matter of fact, due to its
intensity property, the scaled variance of the multiplicity
distribution $\omega[a]$ is widely used to quantify multiplicity
fluctuations in high-energy heavy-ion experiments.
The scaled variance takes the value $\omega[a] = 0$ for $N_a = const.$
and $\omega[a] = 1$ for a Poisson distribution of $N_a$.

\subsection{Strongly intensive quantities}

In nucleus-nucleus collisions the volume of the produced matter (or number of wounded nucleons)
cannot be fixed -- it changes from one event to another.
The quantities, which within the IB-GCE (or WNM) model are independent of $V$ (or $W$) fluctuations are called
\textit{ strongly intensive} quantities~\cite{Gazdzicki:1992ri, Gorenstein:2011vq}.
The ratio of mean multiplicities is
both an intensive and a strongly intensive quantity,
whereas the scaled variance is an intensive but not
strongly intensive quantity.

Strongly intensive quantities $\Delta$ and $\Sigma$ used in this paper are defined as~\cite{Gazdzicki:2013ana}:
\begin{equation}
\label{Eq:siq.delta}
	\Delta[a,b] \equiv \frac{1}{\langle N_b \rangle - \langle N_a\rangle}
	\cdot\Big(\langle N_b \rangle \omega[a] -
	\langle N_a \rangle \omega[b]\Big)
\end{equation}
and
\begin{equation}
\label{Eq:siq.sigma}
	\Sigma[a,b] \equiv  \frac{1}{\langle N_b \rangle + \langle N_a \rangle}
	\cdot \bigg[\langle N_b \rangle \omega[a] +
	\langle N_a \rangle \omega[b]-2\Big(\langle N_a N_b  \rangle -
	\langle N_a \rangle \langle N_b \rangle\Big)\bigg]~,
\end{equation}
where $N_a$ and $N_b$ stand for multiplicities of particles of type $a$ and $b$, respectively.
First and second pure moments,
$\langle N_a \rangle$, $\langle N_b \rangle$, and $\langle N_a^2 \rangle$, $\langle N_b^2 \rangle$
define $\Delta[a,b]$.
In addition, the second mixed moment,
$\langle N_a N_b \rangle$, is needed to calculate $\Sigma[a,b]$.

With the normalization of $\Delta$ and $\Sigma$ used here~\cite{Gazdzicki:2013ana},
the quantities $\Delta[a, b]$ and $\Sigma[a, b]$ are dimensionless and have a common scale required for a
quantitative comparison of fluctuations of different, in general dimensional, extensive quantities.
The values of $\Delta$ and $\Sigma$ are equal to zero in the
absence of event-by-event fluctuations ($N_a=const$., $N_b=const$.)
and equal to one for fluctuations given by the model of
independent particle production (Independent Particle Model)~\cite{Gazdzicki:2013ana}.
The model assumes that particle types are attributed to particles independent of each other. Positive correlations between particle types,
for example \pip and \pim  coming in pairs from $\rho^0$ decays, 
lead to  $\Delta$ and $\Sigma$ values below one. Anti-correlations between particle types, for example due to conservation laws, for example energy conservation leads to anti-correlation of multiplicities of different hadron types, may increase $\Delta$ and $\Sigma$ above one. For detailed discussion see
Refs.~\cite{Begun:2014boa,Gorenstein:2015ria}.

	\section{Identity method}\label{sec:identity}

Experimental measurement of a joint multiplicity distribution of identified hadrons
is challenging. 
Typical tracking detectors, like time projection chambers used by \NASixtyOne,
allow for a precise measurement of momenta of charged particles and  sign of their electric charges.
In order to be able to distinguish between different
particle types (e.g. a particle type $a$ being e$^+$, $\pi^+$, $K^+$ or $p$)
a determination of particle mass is necessary.
This is done indirectly by measuring  for each
particle a value of the specific energy loss \dedx in the tracking detectors,
the distribution of which depends on mass, momentum and charge.
The resolution of \dedx measurements is 
not sufficient for particle-by-particle identification without a radical reduction of considered statistics.
Probabilities to register particles of different types with the same value of \dedx may be comparable.
Consequently, it is impossible to identify particles individually
with reasonable confidence for fluctuation analysis.
The Identity method~\cite{Gazdzicki:2011xz, Gorenstein:2011hr, Rustamov:2012bx, Pruneau:2017fim,  Mackowiak-Pawlowska:2017dma, Pruneau:2018glv}  is a tool to measure moments of
multiplicity distribution of identified particles, which circumvents  the experimental issue of
incomplete particle identification.

The method employs the fitted inclusive \dedx distribution
functions of particles of type $a$, $\rho_{a}(\dedx)$ in momentum bins. Each event has a set
of measured \dedx values corresponding to each track in
the event. For each track in an event the probability w$_{a}$
of being a particle of type $a$ is calculated:
\begin{equation}
\text{w}_{a}= \rho_{a}(\dedx)~/~\rho(\dedx)~,
\end{equation}
where 
\begin{equation}
 \rho(\dedx) = \sum_a \rho_a(\dedx)~.
\end{equation}
Next, an event variable $W_{a}$ (a smeared 
multiplicity of particle $a$ in the event) is defined as:
\begin{equation}
W_{a}=\sum\limits^{N}_{n=1} \text{w}_{a,n}~,
\end{equation}
where $N$ is the number of measured particles in the event. 
The Identity method unfolds  moments of the true
multiplicity distributions from moments of the smeared multiplicity
distribution $P(W_{a})$ using a response matrix calculated from the measured
$\rho_{a}(\dedx)$ distributions~\cite{Gorenstein:2011hr}.

	\section{Experimental setup}\label{sec:setup}

\begin{figure*}
			\centering
			\includegraphics[width=0.8\textwidth]{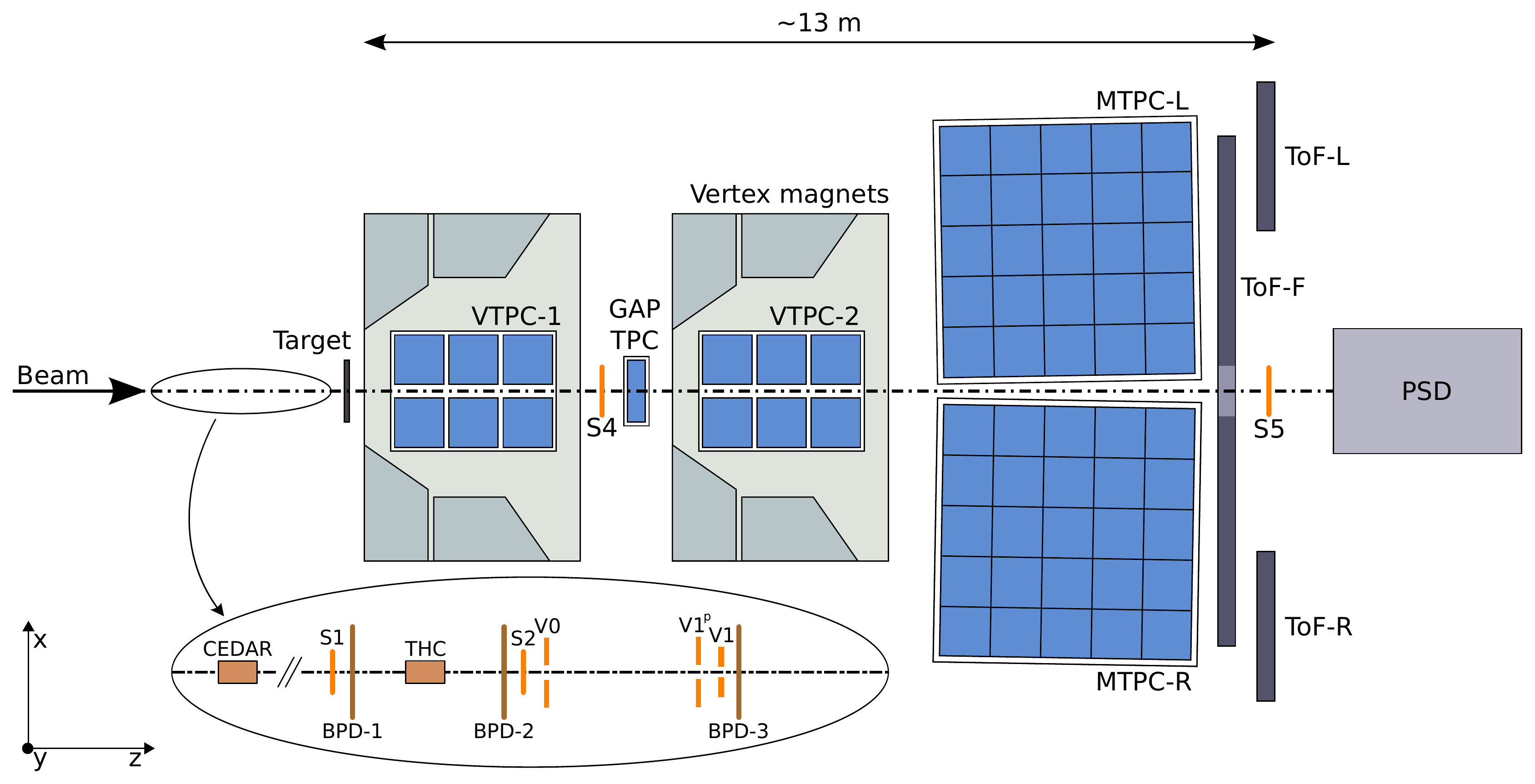}
			\caption[]{
				The schematic layout of
				the \NASixtyOne experiment at the CERN SPS
				(horizontal cut, not
				to scale), see text and Ref.~\cite{Abgrall:2014xwa} for details. The chosen coordinate system is drawn on the lower left: its origin lies in the middle of the VTPC-2, on the beam axis. The nominal beam direction is along the \coordinate{z}-axis. The magnetic field bends charged particle trajectories in the \coordinate{x}-\coordinate{z} (horizontal) plane. Positively charged particles are bent towards the top of the plot. The drift direction in the TPCs is along the \coordinate{y} (vertical) axis.
			}
			\label{fig:detector-setup}
		\end{figure*}

The \NASixtyOne experimental facility~\cite{Abgrall:2014xwa} consists of a large 
acceptance hadron spectrometer located in the H2 beam line of the CERN North Area. 
The schematic layout of the \NASixtyOne detector is shown in Fig.~\ref{fig:detector-setup}. 

The results presented in this paper were obtained using measurement from
the Time Projection Chambers (TPC), the Beam Position Detectors (BPD) and the beam and trigger counters. These detector components as well as the proton beam and the liquid hydrogen target (LHT) are
briefly described below. Further information can be found 
in Refs.~\mbox{\cite{Abgrall:2014xwa, Abgrall:2013qoa, Aduszkiewicz:2017sei}}.

Secondary beams of positively charged hadrons at 31, 40, 80, and
158~\GeVc were produced from 400~\GeVc
protons extracted from the SPS onto a beryllium target.
A selection based on signals from a set of detectors along the H2 beam-line (Cerenkov detectors CEDAR, scintillation
counters S, THC and BPDs (see inset in Fig.~\ref{fig:detector-setup})) allowed to identify
beam protons with a purity of about 99\%. A coincidence of these signals provided the beam trigger $T_{\text{beam}}$. For data taking on p+p interactions a liquid hydrogen target of 20.29~cm
length (2.8\%~interaction length) and 3~cm diameter was placed 88.4~cm 
upstream of the first TPC  (see Fig.~\ref{fig:detector-setup}).
The interaction trigger $T_{\text{int}}$ was provided by the anti-coincidence of
the incoming proton beam and a scintillation counter S4
($T_{\text{int}} = T_{\text{beam}} \wedge\overline{\textrm{S4}}$).
The S4 counter with 2\,cm diameter, was placed
between the magnets  along the  beam trajectory
at about 3.7~m from the target,
see Fig.~\ref{fig:detector-setup}.

The main tracking devices of the spectrometer are four
large volume TPCs. The vertex TPCs (\mbox{VTPC-1} and \mbox{VTPC-2}) are
located in the magnetic fields of two super-conducting dipole
magnets with a maximum combined bending power of 9~Tm which
corresponds to about 1.5~T and 1.1~T fields in the upstream
and downstream magnets, respectively. In order to optimize
the acceptance of the detector, the fields in both magnets were
adjusted proportionally to the beam momentum. Two large main TPCs (\mbox{MTPC-L} and \mbox{MTPC-R})
are positioned downstream of the magnets symmetrically to
the beam line. The fifth small TPC (GAP TPC) is placed
between \mbox{VTPC-1} and \mbox{VTPC-2} directly on the
beam line. It closes the gap between the beam axis and the
sensitive volumes of the other TPCs.
Simultaneous measurements of \dedx and $p_{\mathrm{lab}}$ allow to extract information on
particle mass, which is used to identify charged particles.
Behind the MTPCs there are three Time-of-Flight (ToF) detectors.

	\section{Data reconstruction and simulation}\label{sec:reconstruction}

The event vertex and the produced particle tracks were reconstructed using
the standard \NASixtyOne software~\cite{Abgrall:2013qoa}.
Detector parameters were optimized by a data-based calibration procedure
which also took into account their
time dependence, for details see Refs.~\cite{Abgrall:2013qoa,Abgrall:2008zz}.

A simulation of the \NASixtyOne detector response was used to correct the
reconstructed data. Several Monte Carlo
models were compared with the \NASixtyOne results on p+p, p+C and $\pi$+C
interactions 
~\cite{Abgrall:2013qoa, Aduszkiewicz:2017sei,Abgrall:2011ae,ISVHECRI12_MU}.
Based on these comparisons 
and taking into account continuous support
the \EposLong model~\cite{Werner:2008zza,crmc} was selected for Monte Carlo simulations and
calculation of corrections.
In order to estimate systematic uncertainties simulations were also performed using \VenusLong, which was previously used by the NA49 Collaboration at the CERN SPS energies~\cite{PhysRevC.70.034902,Alt:2006jr}.
Generated and reconstructed tracks were matched based on the number 
of common points along their path. Possible differences due to the different identification procedures followed
in the MC simulations and the real data are addressed in~Ref.~\cite{Aduszkiewicz:2017sei} and Sec.~\ref{sec:sys}.

Since the contribution of elastic events is removed by the event selection (see Sec.~\ref{sec:cuts}),
only inelastic p+p interactions in the hydrogen
of the target cell were simulated
and reconstructed. Thus the MC based corrections (see Sec.~\ref{sec:MCcorr}) 
can be applied only for inelastic events. 

	\section{Event and track selection}\label{sec:cuts} 

The final results presented in this paper refer to identified hadrons produced in inelastic p+p interactions by strong interaction
processes and in electromagnetic decays of produced hadrons. Such hadrons are referred to as \emph{primary} hadrons. The event and track selection cuts described below are selected in
order to minimize unavoidable biases in measured data with respect to the final results.

\subsection{Event selection}

An event was recorded if there was no beam particle detected downstream of  the target (S4 counter in Fig.~\ref{fig:detector-setup}). 
Reconstructed events with off-time beam proton detected within a time window of $\pm$1.5$~\mu$s around the trigger proton were removed. The trajectory of the 
trigger proton was required to be measured in at least three planes out of four of BPD-1 and BPD-2 and in BPD-3. Events without the fitted primary interaction vertex were removed. 
The fitted interaction vertex \coordinate{z}-coordinated was requested to be within $\pm$20~cm from 
the LHT target center.
Events with a single positively charged track with absolute momentum close to the beam momentum (for details see Ref.~\cite{Abgrall:2013qoa}) are removed in order to eliminate remaining elastic scattering reactions. This requirement removes less then 2$\%$ of all 
events in data at 31-80 \GeVc beam momenta. 
The same cut applied to the \Epos simulated and reconstructed data removes less then 5$\%$ of all inelastic events. The cut was not applied to the 158~\GeVc data-set.
The fraction of inelastic events removed by the event selection cuts estimated based on the \Epos simulation ranges between 15$\%$ at 31~\GeVc and 22$\%$ at 158~\GeVc.

\subsection{Track selection}

In order to select good quality tracks of primary charged hadrons and to reduce the contamination
of tracks from secondary interactions, weak decays and off-time interactions, the following track selection criteria
were applied:
\begin {enumerate}[(i)]
\item Track momentum fit at the interaction vertex should have converged.
\item Total number of reconstructed points used to fit the track trajectory  should be greater than 30.
\item Sum of the number of reconstructed points in VTPC-1 and VTPC-2 should be greater than 15 or the number of reconstructed points in the GAP TPC should be greater than 4.
\item Distance between the track extrapolated to the interaction plane and the interaction point (track impact parameter)
should be smaller than 4~\cm in the horizontal (bending) plane and 2~\cm in the vertical. (drift) plane,
\item Total number of points used to obtain track \dedx  should be greater than 30.
\item A track is measured in the high efficiency region of the detector and it should lie in the region where \dedx measurements are available (see Ref.~\cite{na61ParticlePopulationMatrix}.
This defines the analysis acceptance
given in Ref.~\cite{na61ParticlePopulationMatrix} in a form of three dimensional maps
in particle momentum space.
Examples of the analysis acceptance are shown in Fig.~\ref{fig:acc-ptflu}.

\end {enumerate}

The event and track statistics after applying the selection criteria
are summarized in Table~\ref{tbl:data-sets}. 
\begin{table}
	\vspace{0.2cm}
	\caption {
		Statistics of accepted events as well as 
		number of accepted positively and negatively charged tracks for data analysed in the paper. 
	}
	\begin{center}
		\begin{tabular}{|c|c|c|c|}
			\hline
			Beam momentum & \# events & \# positively 	& \# negatively 	
			 \cr
			[\GeVc] &  & charged tracks & charged tracks	
			\cr 
			\hline
			31 &  819710 & 530971	&	132187	
			\cr
			40 &  2641412 & 2071490 	&	675258	
			\cr
			80 &  1531849 & 2061069 &	1020267	
			\cr
			158 & 1587680 & 3243819	&	1980037	
			\cr \hline
		\end{tabular}
	\end{center}
	\label{tbl:data-sets}
\end{table}

	\section{Identity analysis}\label{sec:identity_analysis}

In order to calculate moments of multiplicity distributions of identified hadrons corrected
for incomplete particle identification the analysis was performed
using the Identity method. The analysis consists of three steps:
\begin{enumerate}[(i)]
	\item parametrization of inclusive \dedx spectra,
	\item calculation of smeared multiplicity distributions and their moments,
	\item correction of smeared moments for incomplete particle identification using the \dedx response matrix.
\end{enumerate}
The Identity analysis steps are briefly described below and in App.~\ref{sec:app2}.

\begin{figure}[!ht]
	\begin{center}
		\includegraphics[width=0.9\textwidth]{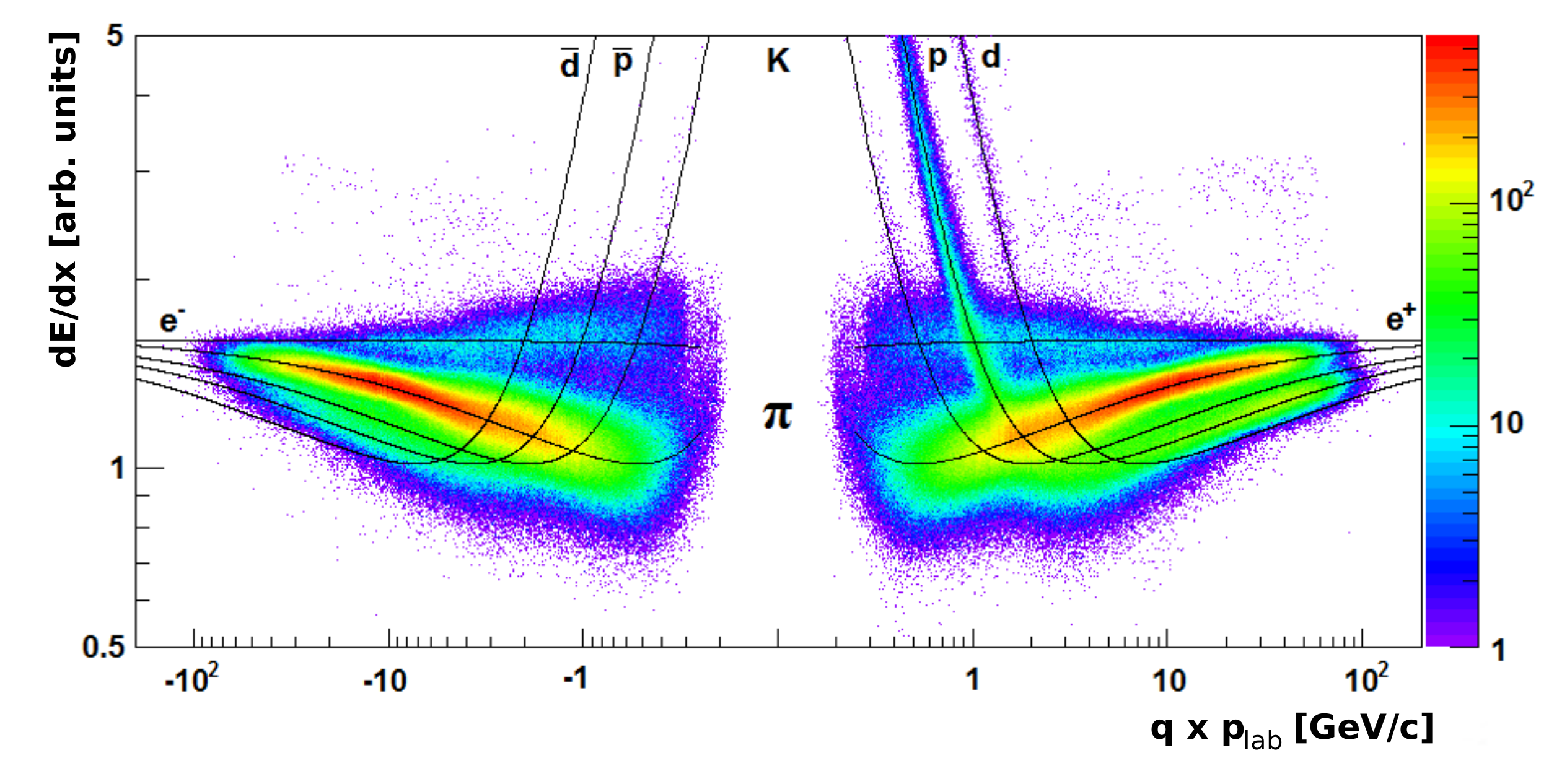}
	\end{center}
	\caption{
		Example distribution of charged particles in the \dedx -- $q \cdot p_{\mathrm{lab}}$ plane
		in p+p interactions at 80~\GeVc.
		Expectations for the dependence of the mean \dedx on $p_{\mathrm{lab}}$ for the considered particle
		types are shown by the curves calculated based on the Bethe-Bloch function.
	    }
	\label{fig:dedx}
\end{figure}

For each particle its specific energy loss
\dedx is calculated as the truncated mean (smallest 50\%)
of cluster charges measured along the track trajectory.
As an example, \dedx measured in p+p interactions at 80~\GeVc, for positively and negatively charged particles,
as a function of $q \cdot p_{\mathrm{lab}}$ is presented in Fig.~\ref{fig:dedx}.
The expected mean values of \dedx for different particle types are shown by the Bethe-Bloch curves.

The parametrization of \dedx spectra of $e^{+}$, e$^{-}$, $\pi^{+}$, $\pi^{-}$, $K^{+}$, $K^{-}$,
\emph{p}, and $\bar{\emph{p}}$ were obtained by fitting the \dedx distributions separately for positively
and negatively charged particles in bins of  $p_{\mathrm{lab}}$ and transverse momentum $\pt$.

The fitted function was defined as a sum of 
the \dedx distributions of $e^{+}$, $\pi^{+}$, $K^{+}$,
and \emph{p} for positively charged particles. The sum of the contributions
of the corresponding antiparticles was used for negatively charged particles.

The details of this fitting procedure
can be found in 
Refs.~\cite{Aduszkiewicz:2017sei,vanLeeuwen:2003ru,note_MvL}. 
In contrast to the spectra analysis~\cite{Aduszkiewicz:2017sei} separate fits were performed in order to extend acceptance by adding particles with negative $p_\coordinate{x}/q$, where $p_\coordinate{x}$ is \coordinate{x}-component of the particle total momentum in the laboratory system.
Systematic uncertainties arising from
the fitting procedure are estimated in Sec.~\ref{sec:other}.
In order to ensure similar particle numbers in each bin, 20 logarithmic bins
were chosen in $p_{\mathrm{lab}}$ in the  range $1-100$~\GeVc.
Furthermore, the data were binned in 20 equal \pt intervals in the range $0-2$~\GeVc.

\begin{figure}[!ht]
	\begin{center}
		\includegraphics[width=0.45\textwidth]{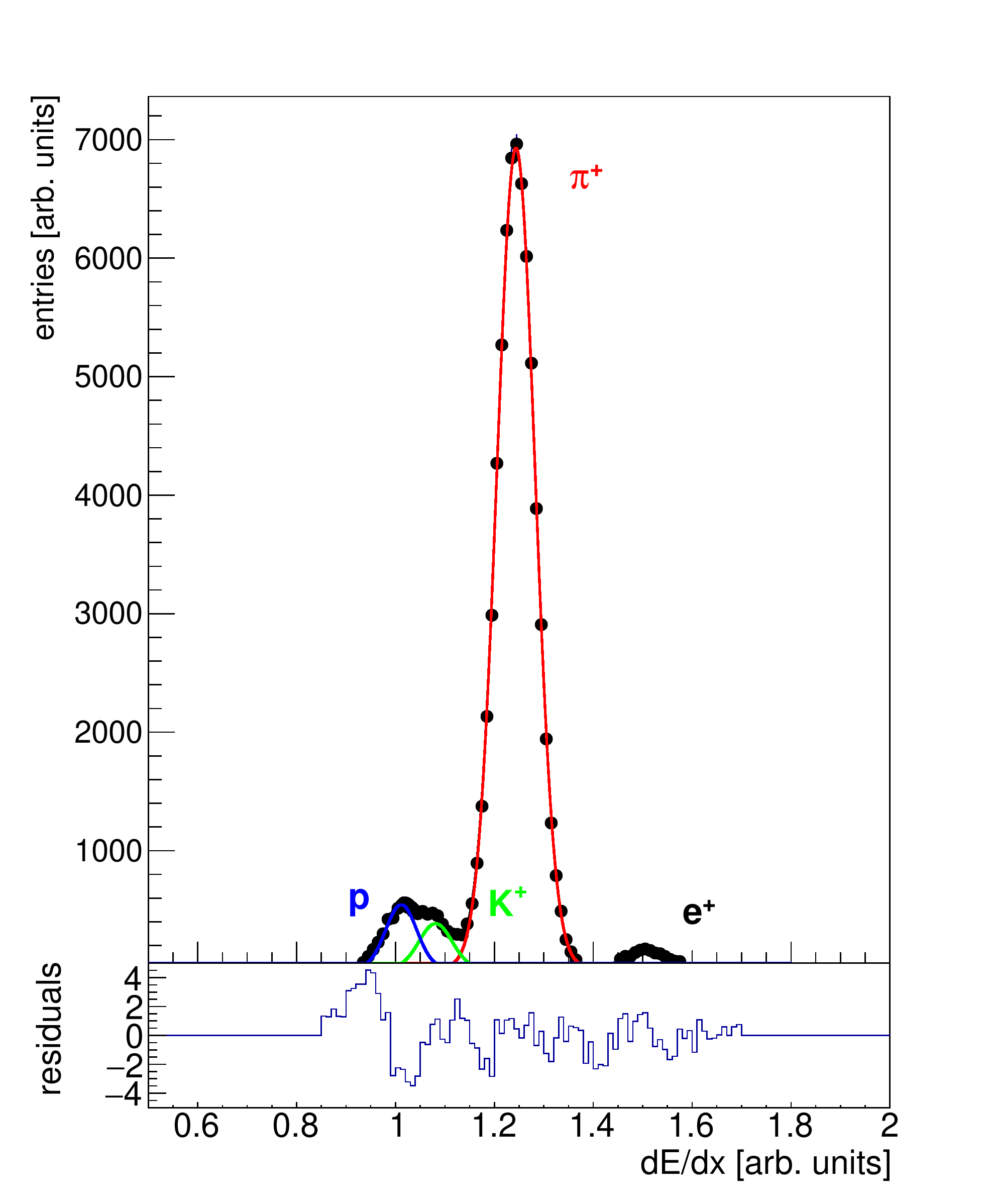}
		\includegraphics[width=0.45\textwidth]{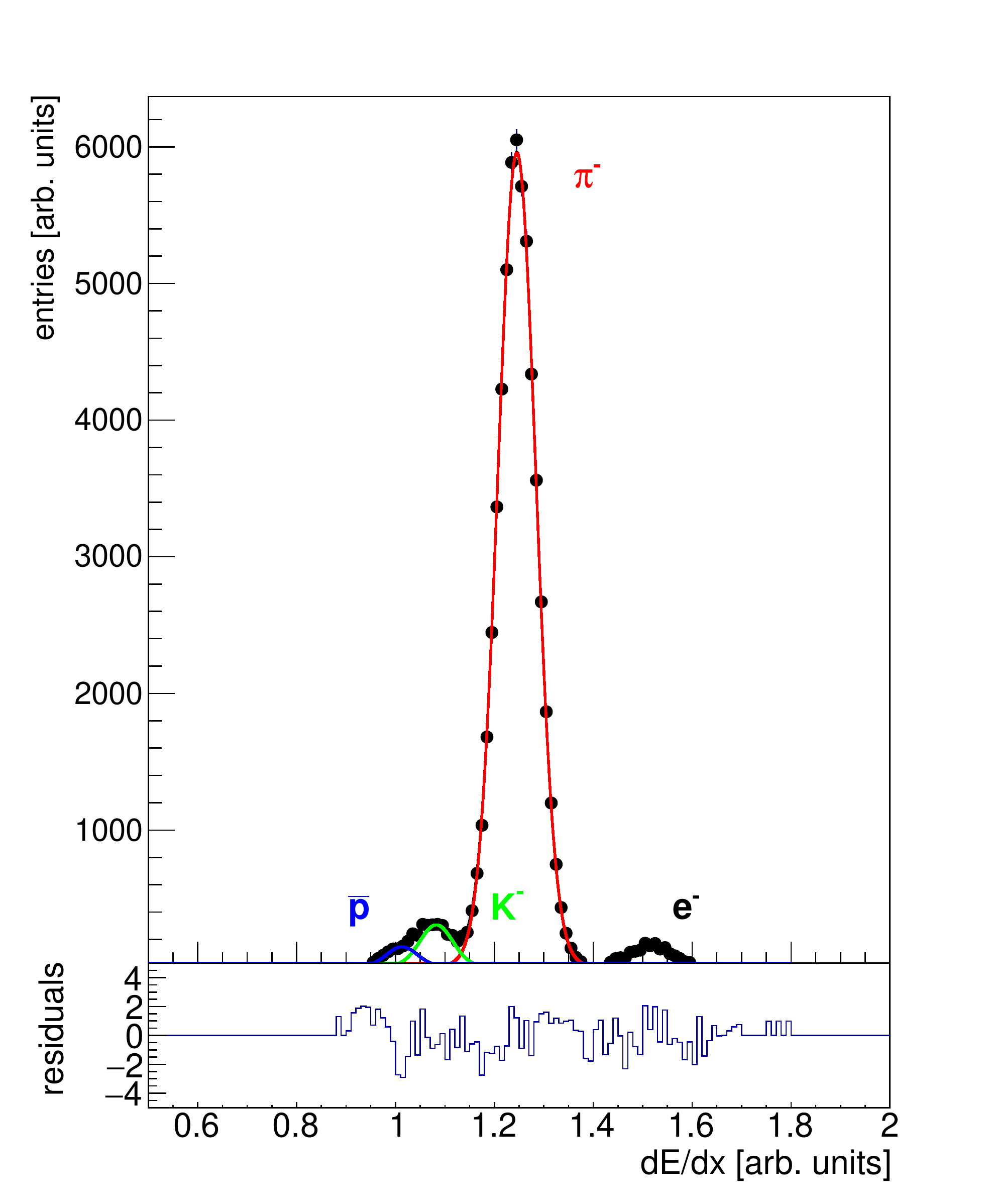}\\
		\includegraphics[width=0.45\textwidth]{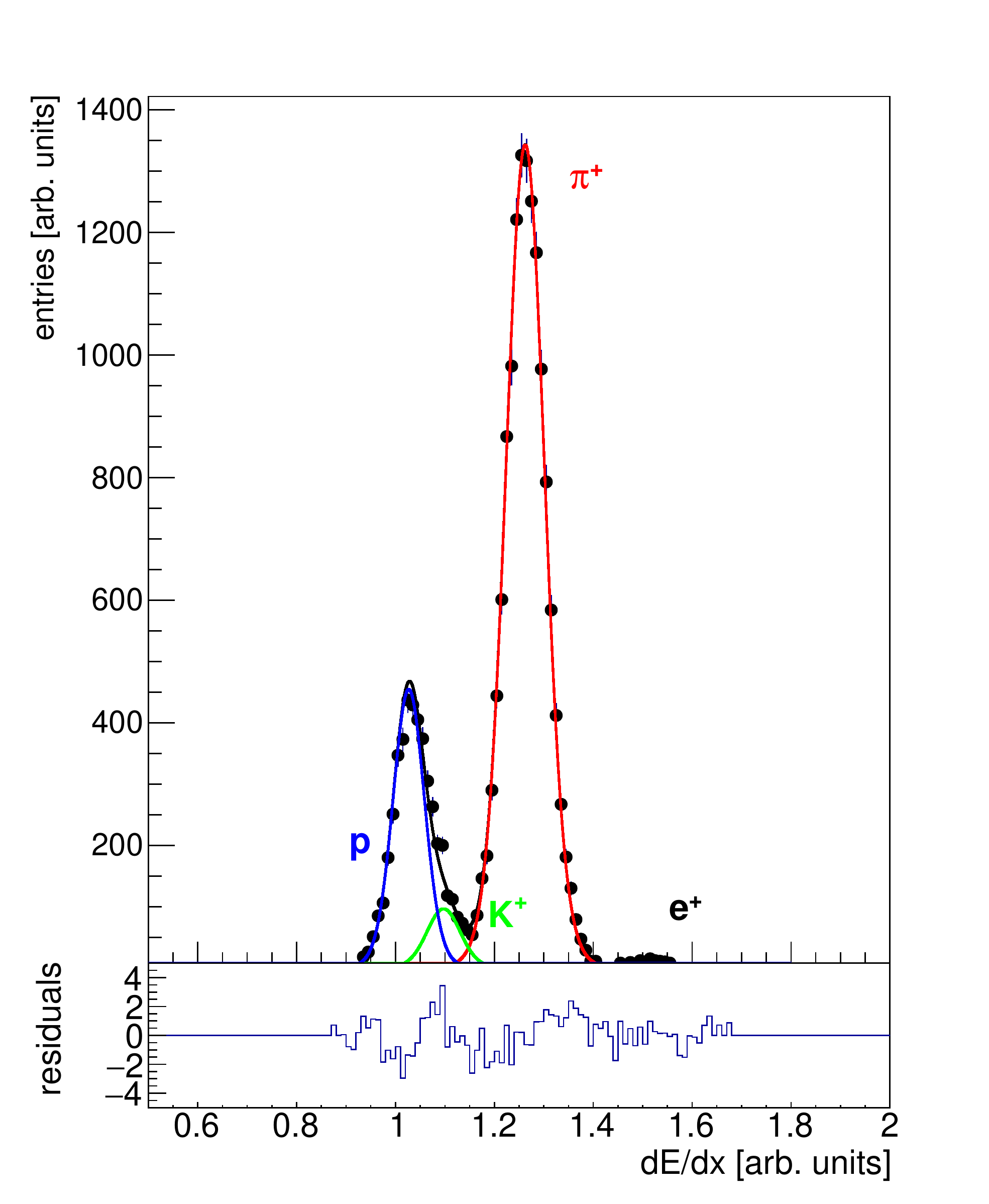}
		\includegraphics[width=0.45\textwidth]{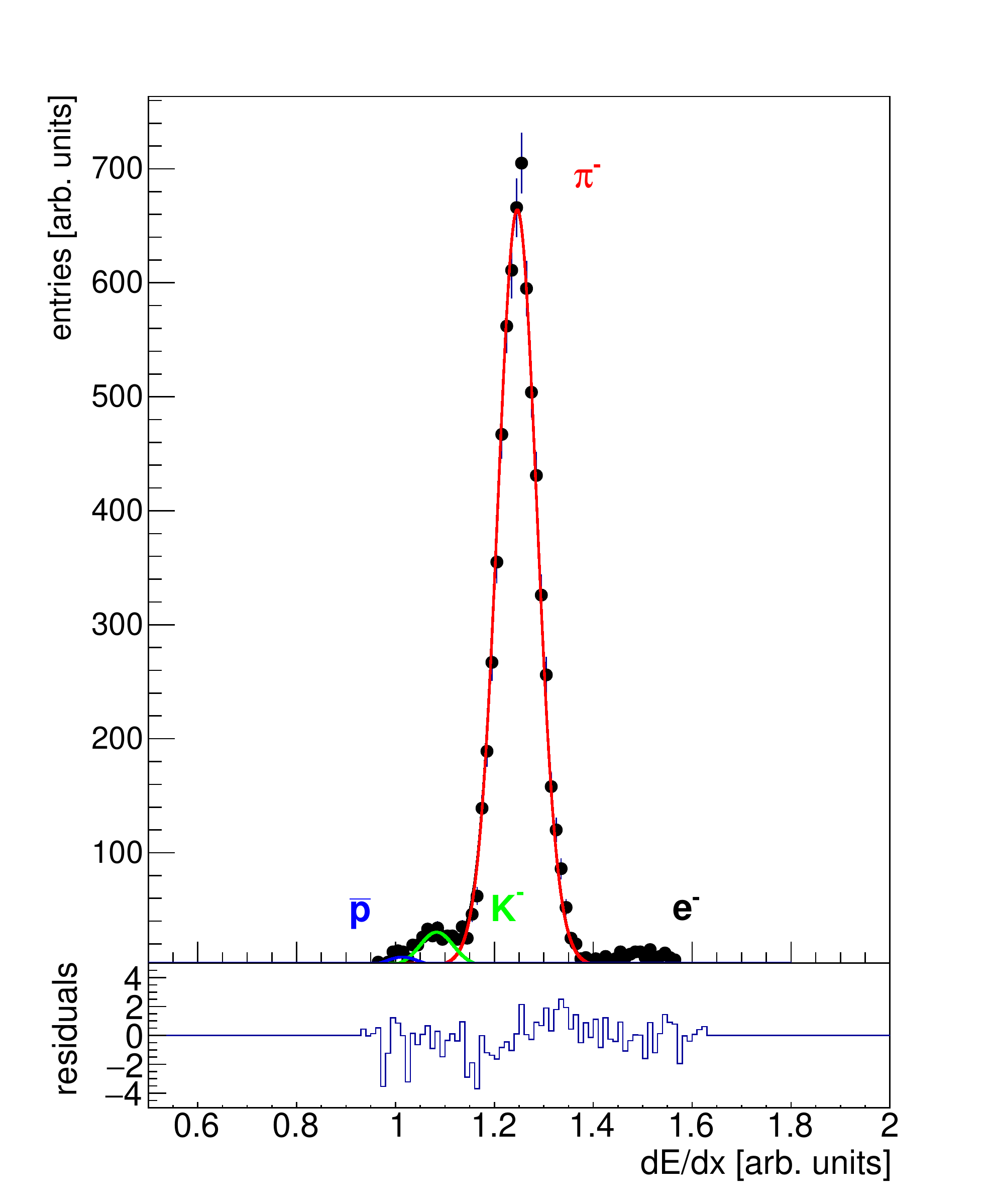}
	\end{center}
	\caption{
		The \dedx distributions for positively (\textit{left})
		and negatively (\textit{right}) charged
		particles in the bin 5.46 $< p_{\mathrm{lab}} \leq$ 6.95~\GeVc and
		0.1 $< \pt \leq$ 0.2~\GeVc produced in p+p interactions at 158~\GeVc (\textit{top}) and 31~\GeVc (\textit{bottom}).
		The fit by a sum of contributions from different particle types is shown by black lines.
		The corresponding residuals (the difference between the data and fit divided
		by the statistical uncertainty of the data) is shown in the bottom of the plots.
     	}
	\label{fig:exfit}
\end{figure}

The \dedx spectrum for a given particle type was parametrized by the sum of 
\emph{asymmetric Gaussians} with
widths $\sigma_{a,l}$ depending on the particle type $a$ and the number of points $l$ measured in the TPCs.
The peak position of the \dedx distribution
for particle type $a$ is denoted as $x_{a}$.
The contribution of a reconstructed particle track to the fit function reads:

\begin{equation}
\rho(x)=\sum_{a}\rho_{a}(x)=\sum\limits_{a=\pi,\textrm{p},\textrm{K},\textrm{e}} Y_{a} \frac{1}{\sum\limits_{l} n_{l}} \sum\limits_{l} \frac{n_{l}}{\sqrt{2\pi}\sigma_{l}}\exp \left [-\frac{1}{2}\left (\frac{x-x_{a}}{(1\pm\delta)\sigma_{l}} \right ) ^{2} \right ]~,
\label{Eq:AsymGaus}
\end{equation}

where $x$ is the \dedx of the particle, $n_{l}$ is the number of tracks with number of points $l$
and $Y_{a}$ is the amplitude of the contribution of particles of type $a$.
The second sum is the weighted average of the line-shapes from the different numbers of measured points
(proportional to track-length) in the sample. The quantity $\sigma_{l}$ is written as:

\begin{equation}
\sigma_{l}=\sigma_{0}\left ( \frac{x_{i}}{x_{\pi}}\right )^{0.625} \Big/ \sqrt{l}~,
\label{Eq:sigma}
\end{equation}
where the width parameter $\sigma_{0}$ is assumed to be common for all particle types and bins.
A $1/\sqrt{l}$ dependence on number of points is assumed following Ref.~\cite{Marco_fit}. The asymmetry parameter  $\delta$ is introduced in order
to take into account a possible small asymmetry of the truncated mean distribution resulting
from a strong asymmetry of the Landau energy loss distribution. Examples of fits for p+p interactions at 31 and 158~\GeVc are shown in Fig.~\ref{fig:exfit}.

\begin{figure*}
	\centering
	\includegraphics[width=0.4\textwidth]{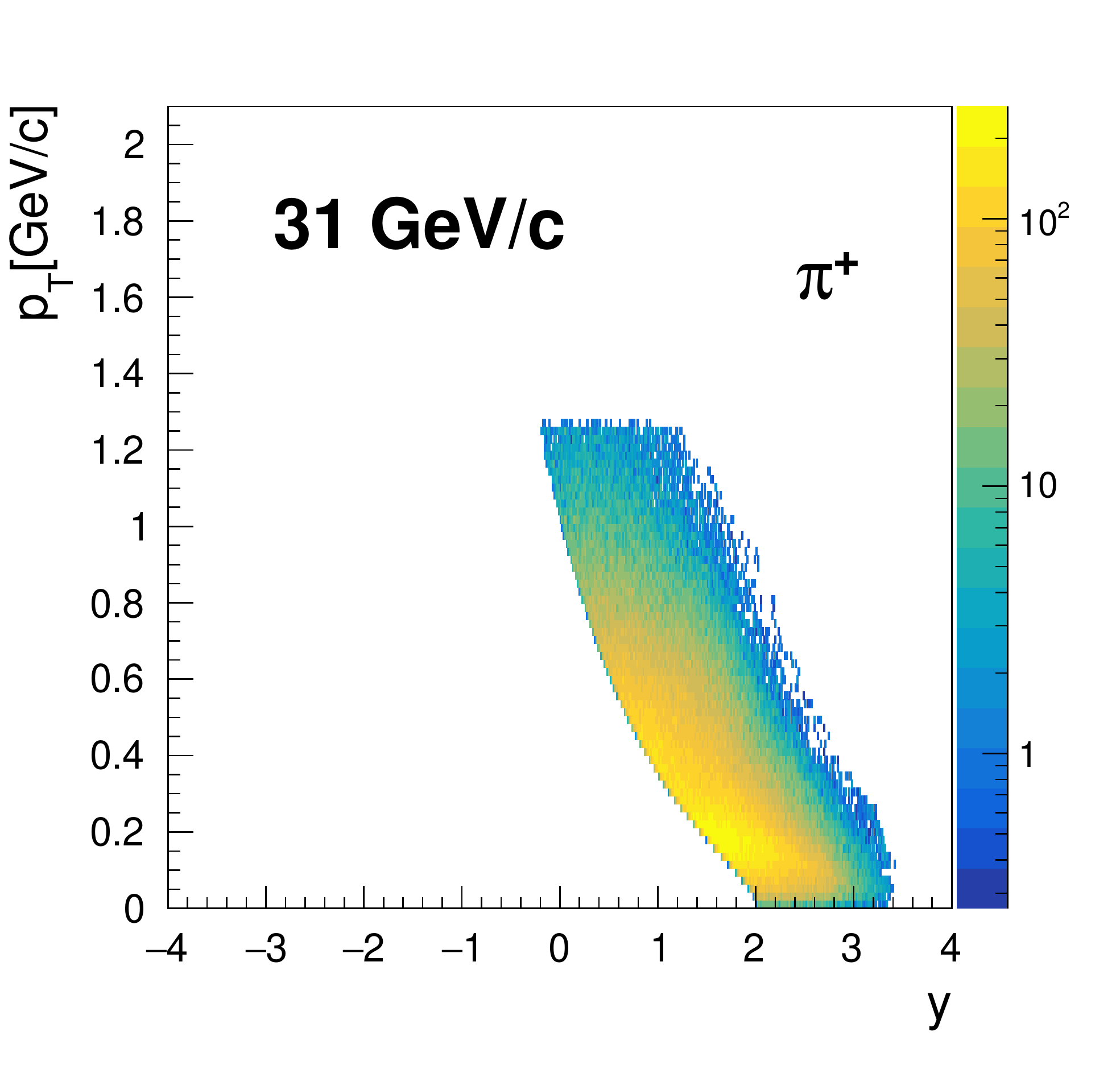}
	\quad
	\includegraphics[width=0.4\textwidth]{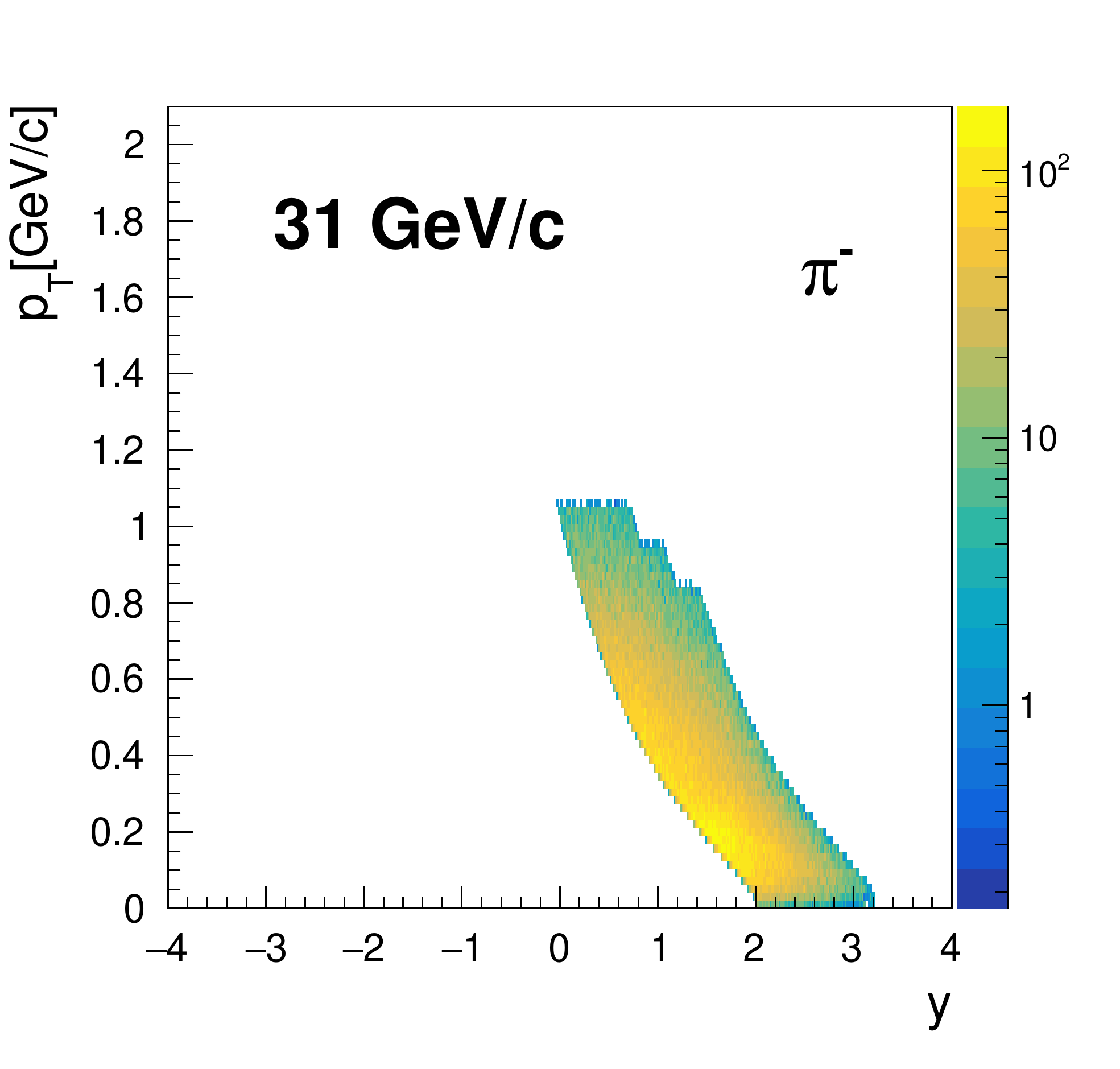}
	\\
	\includegraphics[width=0.4\textwidth]{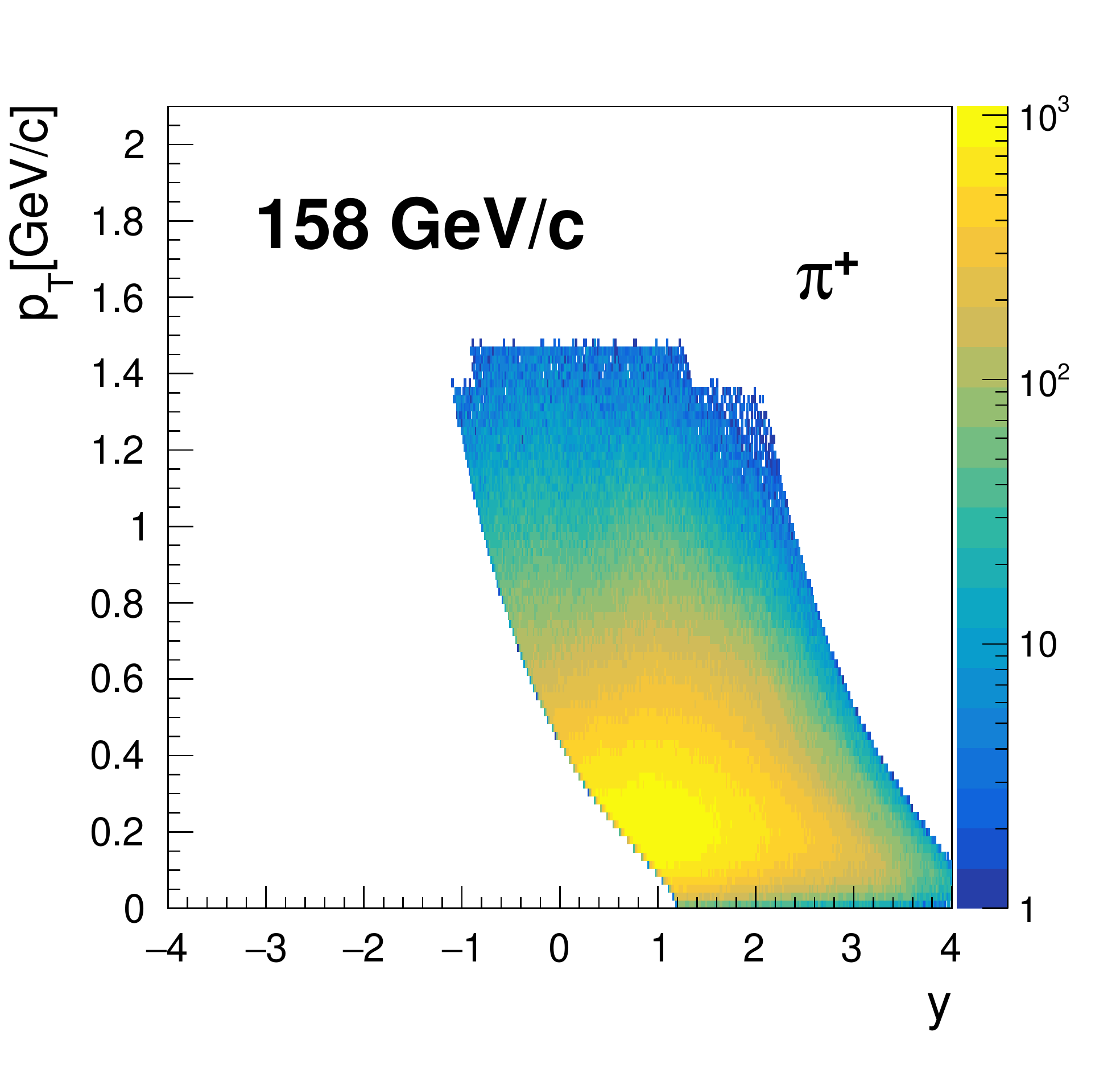}
	\quad
	\includegraphics[width=0.4\textwidth]{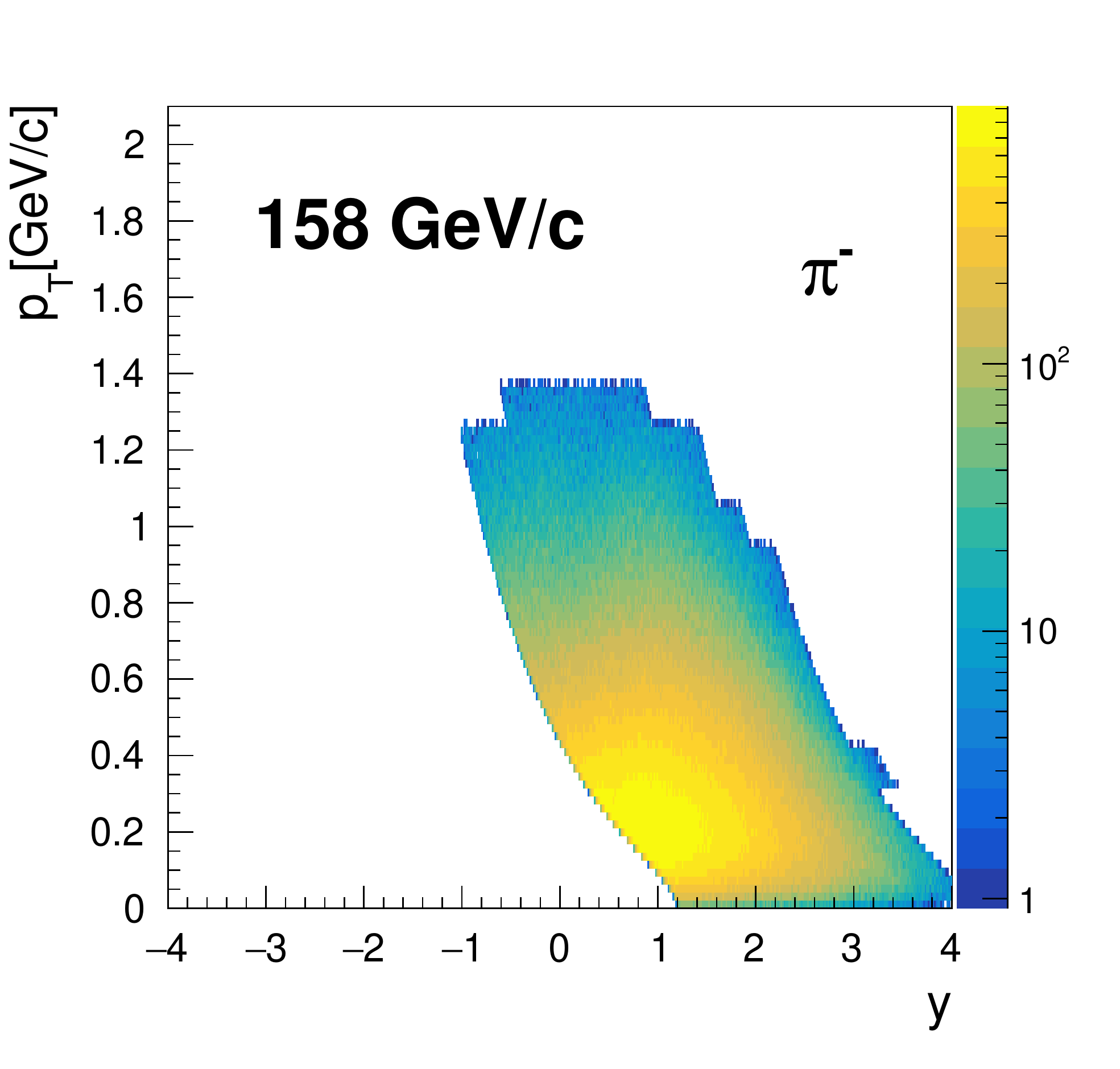}
\caption[]{
     	Distributions of particles selected for the analysis in transverse momentum \pt and
     	 rapidity $y$ calculated in the collision center-of-mass reference system assuming the pion mass. The two upper plots are for 31~\GeVc and the two lower plots for 158~\GeVc. The irregular edges of the
        distributions reflect the boundaries of the $p_{\mathrm{lab}}$, \pt bins used in the \dedx analysis.
	}
	\label{fig:acc-ptflu}
\end{figure*}

In order to ensure good fit quality, only bins with number of tracks greater than 500 were used for further analysis.
The Bethe-Bloch curves for different particle types cross each other at low values of the total momentum.
Thus, the proposed technique is not sufficient for particle identification at low $p_{\mathrm{lab}}$ and
bins with $p_{\mathrm{lab}}<$~4.3~\GeVc  were excluded from this analysis based solely on \dedx.
The requirement of at least 500 tracks with good quality \dedx measurement
in each $p_{\mathrm{lab}}$, \pt bin reduces
the acceptance available for the analysis. Due to different multiplicities the acceptance
is different for positively and negatively charged particles.
Moreover, it also changes with beam momentum. Thus, the largest acceptance was
found for positively charged hadrons at 158~\GeVc and the smallest at 31~\GeVc for negatively charged hadrons.
The acceptance used in this analysis is given separately for
negatively and positively charged particles by a
set of publicly available acceptance tables~\cite{na61ParticlePopulationMatrix}.
The corresponding rapidity and transverse momentum acceptances at 31 and 158~\GeVc
are shown in Fig.~\ref{fig:acc-ptflu}.

The parametrization of inclusive \dedx spectra of identified particles is first used to calculate probabilities w$_a$ and, then, W$_a$. The first moments of the multiplicity distributions for complete particle identification,
$\langle N_a \rangle$ are equal
to the corresponding first moments of the smeared distributions:
\begin{equation}
	\langle N_a \rangle = \langle \textrm{W}_a \rangle~.
\label{Eq:first}
\end{equation}
Second  moments of the multiplicity distributions of identified hadrons are obtained
by solving sets of linear equations which relate them to the corresponding smeared moments.
The parameters of the equations are calculated using   the \dedx densities of identified particles obtained from the fits to the experimental data. Details can be found in App.~\ref{sec:app2}.

The Identity method was quantitatively tested by numerous simulations, see for example
Refs.~\cite{Rustamov:2012bx,Mackowiak-Pawlowska:2017dma}.

	\section{Corrections and uncertainties}
\label{sec:other} 

This section briefly describes the corrections for biases and presents methods to calculate statistical and
systematic uncertainties.

\subsection{Corrections for event and track losses and contribution of unwanted tracks}
\label{sec:MCcorr}

The first and second moments of multiplicity distributions corrected for incomplete particle identification were also corrected for:
\begin{enumerate}[(i)]
	\item loss of inelastic events due to the on-line and off-line event selection,
	\item loss of particles due to the detector inefficiency and track selection,
	\item contribution of particles from weak decays and secondary interactions (feed-down).
\end{enumerate}				 

A simulation of the \NASixtyOne detector response was used to 
correct the data for the above mentioned biases.    
Corrections were calculated for moments of identified hadron multiplicity distributions. 
Events simulated with the \Epos model were reconstructed with the standard \NASixtyOne software
as described in Sec.~\ref{sec:reconstruction}. 
The multiplicative correction factors $C_{a}^{(k)}$ and $C_{ab}$, where $a$ and $b$ denote the particle type ($a,b=\pi^{+/-}, K^{+/-}, p, \bar{p}, e^{+/-}$; and $a\neq b$), are defined as:
\begin{equation}
C_{a}^{(k)}=\frac{(N_{a}^{k})_{gen}^{MC}}{(N_{a}^{k})_{sel}^{MC}}, \quad C_{ab}
= \frac{(N_{ab})_{gen}^{MC}}{(N_{ab})_{sel}^{MC}}~,
\end{equation}
where:
\begin{enumerate}[(i)]
	\item $(N_a^{k})_{gen}^{MC}$ -- moment $k$ of particle type $a$ generated by the model,
	\item $(N_a^{k})_{sel}^{MC}$ -- moment $k$ of particle type $a$ generated by the model with the detector response simulation, reconstruction and
	selection,
	\item $(N_{ab})_{gen/sel}^{MC}$ -- mixed second moment of particle types $a$ and $b$ 
	generated by the model ($gen$) and with the detector response simulation, reconstruction and
	selection ($sel$).
\end{enumerate}
\begin{figure}[!ht]
	\begin{center}
		\includegraphics[width=0.8\textwidth]{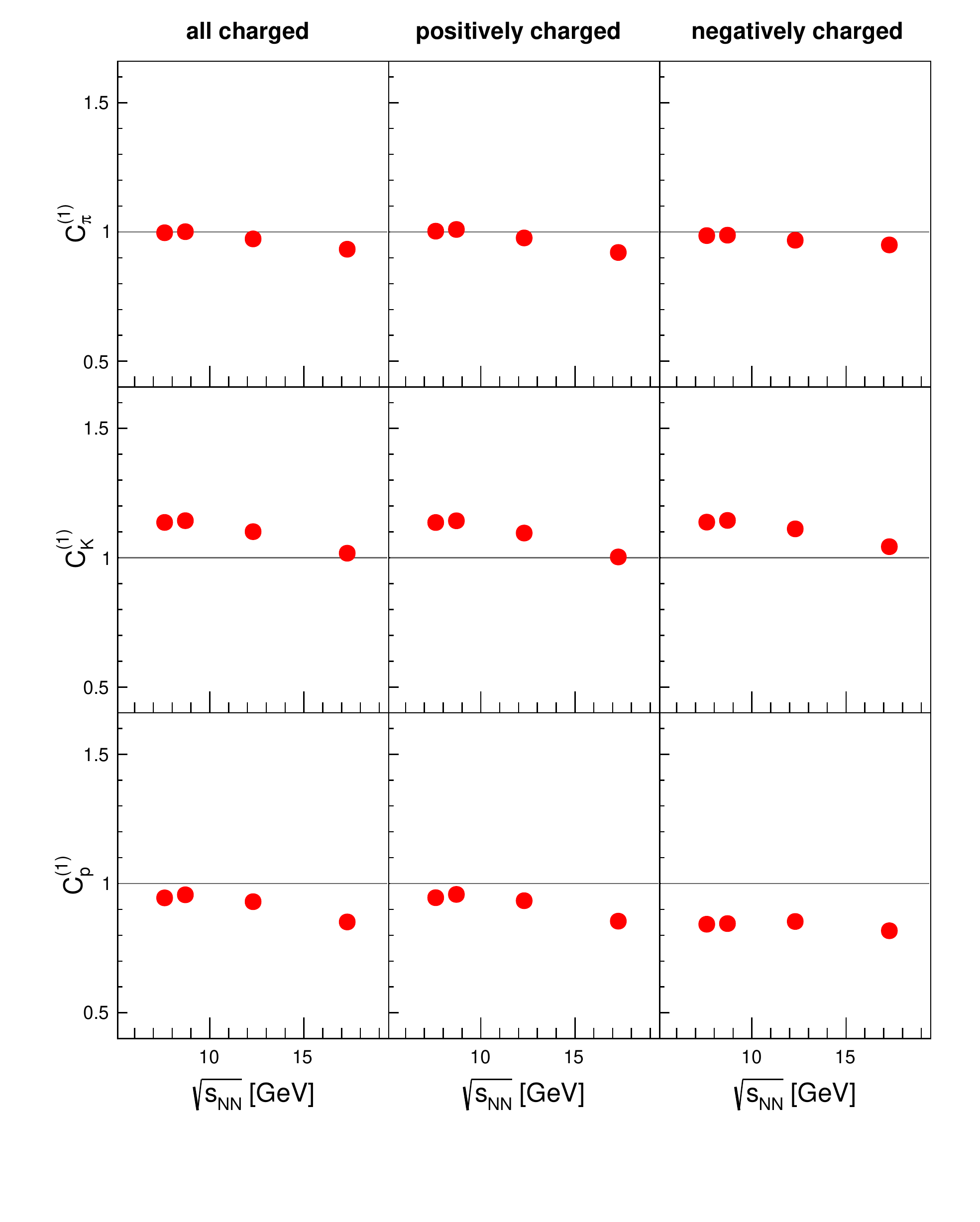}
	\end{center}
	\caption{      
		Energy dependence of correction factor $C_{a}^{(1)}$ for all charged, positively and negatively charged pions, kaons and protons.
	}
	\label{fig:CF1st}
\end{figure}
\begin{figure}[!ht]
	\begin{center}
		\includegraphics[width=0.8\textwidth]{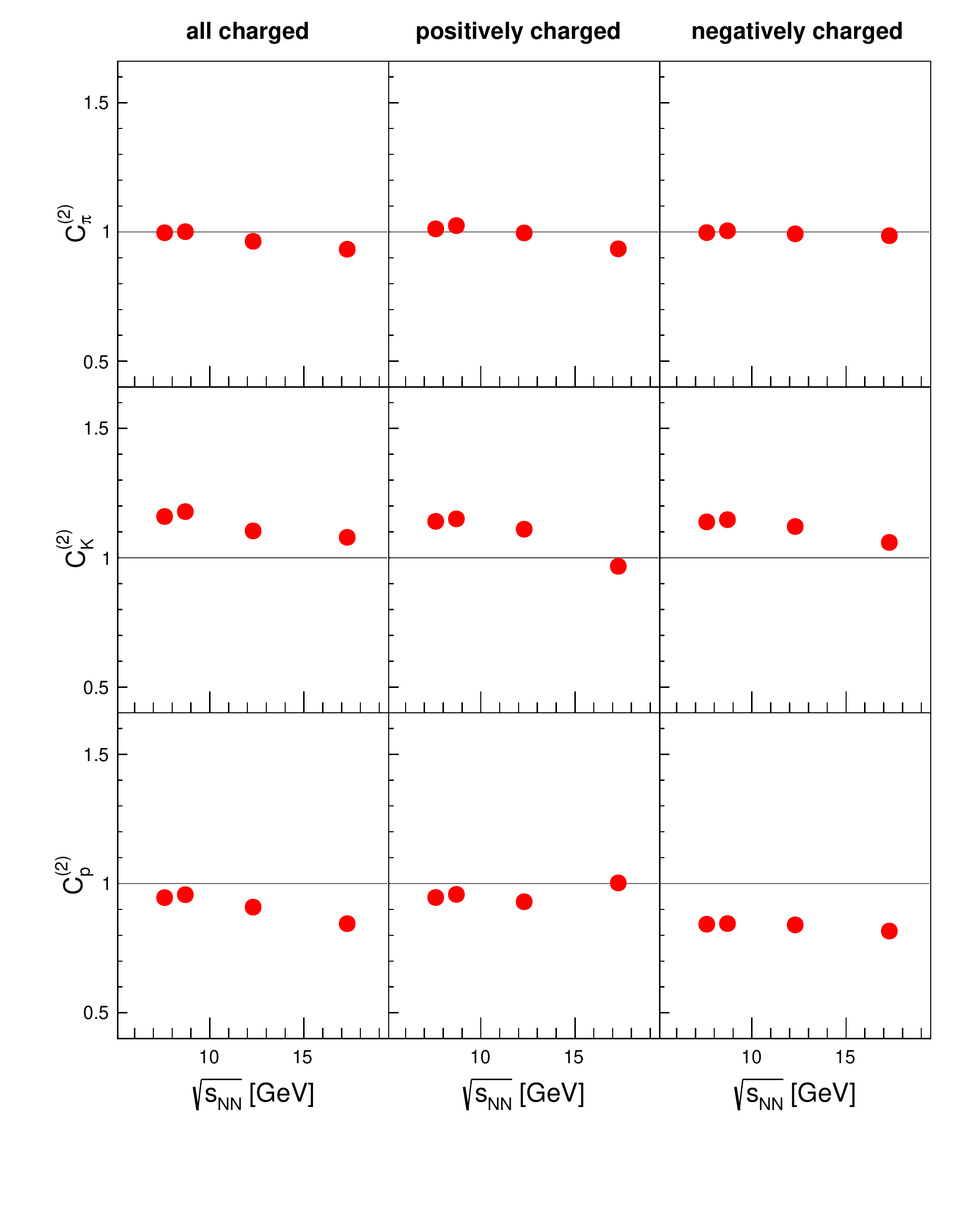}
	\end{center}
	\caption{     
		Energy dependence of correction factor $C_{a}^{(2)}$ for all charged, positively and negatively charged pions, kaons and protons.
	}
	\label{fig:CF2nd}
\end{figure}
\begin{figure}[!ht]
	\begin{center}
		\includegraphics[width=0.8\textwidth]{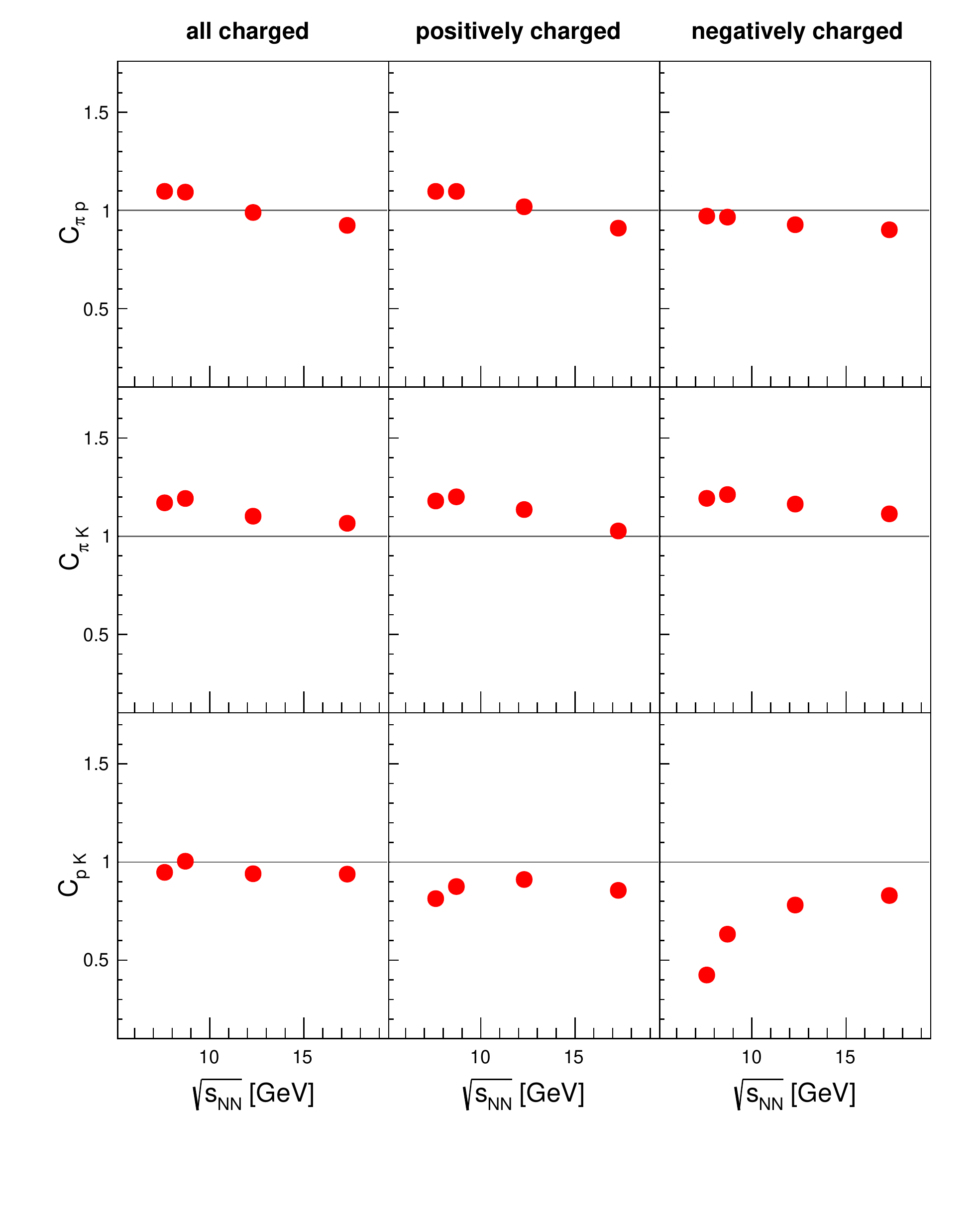}
	\end{center}
	\caption{      
		Energy dependence of correction factor $C_{ab}$ for all charged, positively and negatively charged combinations of pions, protons and kaons.
	}
	\label{fig:CFmixed}
\end{figure}
This way of implementing the corrections was tested using simulations based on the \Epos and \Venus models, for details
see Sec.~\ref{sec:sysmodel}.
Multi-dimensional unfolding of distributions which would be 
the best approach for correcting experimental 
biases is too complex to be implemented~\cite{Adye:2011gm}.
This is why the Identity method - the unfolding of moments -  
was used to correct for the main bias -- the incomplete particle identification.
Then the unfolded moments were corrected for remaining biases.

The correction factors for first, second and mixed moments of identified hadrons are shown in Figs.~\ref{fig:CF1st},~\ref{fig:CF2nd} and~\ref{fig:CFmixed}.
Note that a single particle reconstruction inefficiency is typically lower than 10\%, 
whereas the feed-down contribution is about 10\%~\cite{Abgrall:2013qoa,Aduszkiewicz:2017sei}.

\subsection{Statistical uncertainties}
\label{sec:stat}

The \textit{ sub-sample method} was used to calculate statistical uncertainties of final results. 
All selected events were grouped into $M=30$ non-overlapping sub-samples of events.
Then a given fluctuation measure $Q$ (for example $\Sigma[\pi^+,p]$) was calculated for each
sub-sample separately, and the variance of
its distribution, $Var[Q]$, was obtained. 
The statistical uncertainty of $Q$ for all selected events was estimated as
$\sqrt{Var[Q]/M}$. 
The $\dedx$ parametrization requires a minimum number of tracks in a given momentum bin,
thus the acceptance in which the \dedx parametrization can be obtained is larger for all selected events
than for sub-samples of events. In order to have the maximum acceptance the same \dedx parametrization  
obtained using all events was used in the sub-sample analysis.
It was checked using the \textit{bootstrap method}~\cite{efron, moore} that the above approximation
leads only to a small underestimation of statistical uncertainties.

\clearpage

\subsection{Systematic uncertainties}
\label{sec:sys}

Systematic uncertainties originate from imperfectness of the detector response and systematic uncertainties in the modelling of physics processes implemented in the models. The total systematic uncertainties were calculated by adding detector-related (see Sec.~\ref{sec:sys_det}) 
and model-related (see Sec.~\ref{sec:sysmodel})  contributions in quadrature. 

\subsubsection{Detector related effects}
\label{sec:sys_det}

These uncertainties were studied by applying standard (see Sec.~\ref{sec:cuts}) and loose cuts defined by:
\begin{enumerate}[(i)]
    \item reducing the rejection window for events with off-time beam particle to <~0.5~$\mu$s,
    \item  relaxing the requirement on the $z$ position of the main vertex (to exclude off-target events and inelastic p+p interactions),
    \item reducing the requirement of the minimum number of measured points in the detector to 20,
    \item loosening the constraint on the distance of the track extrapolated back to
the target plane and the main vertex from 4 to 8~cm and from 2 to 4~cm in the $x$ and $y$ directions.
\end{enumerate} 
For each choice the complete analysis was repeated including the \dedx fitting and recalculating the corrections. 
Observed differences between results for the standard and loose cuts  are related to imperfectness of the 
reconstruction procedure and to the acceptance of events
with additional tracks from off-time particles.
The corresponding systematic uncertainty was calculated as the difference between the results
for the standard and loose cuts.

An additional possible source of uncertainty is imperfectness of the \dedx parametrization.
Here the largest uncertainty comes from
uncertainties of the parameters of the kaon \dedx distribution.
The kaon distribution significantly overlaps with the proton and pion distributions.
In the most difficult low momentum range the \dedx fits were cross-checked using the 
time-of-flight information and found to be in agreement at the level of single 
particle spectra (see Ref.~\cite{Aduszkiewicz:2017sei}).

In this analysis, as it considers second order moments, two additional tests were performed. 
First, fits of \dedx distributions with fixed asymmetry parameter and without any constraint on asymmetry 
were used to estimate the resulting possible biases of fluctuation measures. The change of the results is 
below 10\% for most quantities. Larger relative differences appear only for quantities close to 0. 
A second test was performed to validate fit stability. 
The value of \dedx for each reconstructed track in the Monte-Carlo simulation was generated using the parametrization of the \dedx response function fitted to the data.
Next,
\dedx fits were performed on reconstructed \Epos simulated events. Intensive and strongly intensive quantities 
were obtained the same way as in the data and compared to the values obtained in the model on the generated level. 
The change of the results is below 10\% for most quantities and, for almost all, it is 
within or comparable to the systematic uncertainty. The only exceptions are the scaled variance 
of protons at 158~\GeVc~(10$\%$ which normally is 5$\%$) and pions at 31~\GeVc~(15$\%$ which normally is 8$\%$) as well as $\Delta$ of pions and protons at 158~\GeVc~
($17\%$ compared to $11\%$). 

\subsubsection{Model-related effects}
\label{sec:sysmodel}
Systematic uncertainty originating from imperfectness of the \Epos model used to calculate corrections in
describing \pp interactions is discussed here.
The uncertainty was estimated using simulations performed within the \Epos and \Venus models.
The simulated \Epos data  were corrected using corrections obtained based on the \Venus model and compared to the unbiased \Epos results. Then the same procedure was repeated swapping 
\Epos and \Venus.
In both cases the correction improves agreement between the
obtained results and the true ones.
The differences between the unbiased and simulated-corrected results
were  added to the systematic uncertainty. 
They are on average about 20\% (\Epos data) and 25\% (\Venus data) of the total systematic uncertainty.
Note that the  models show  similar
agreement with results on p+p interactions at the CERN SPS energies.

	\section{Results, discussion and comparison with models}
\label{sec:results}

In this section final experimental results are presented and discussed as well as compared with predictions of selected string-hadronic models.

\begin{table}
\centering
\begin{tabular}{|c|c|c|c|c|c|c|}
\hline
Beam momentum [\GeVc]	&	$\langle \pi\rangle_{acc}$	 &	$\langle \pi\rangle$&	$\langle K\rangle_{acc}$ &	$\langle K\rangle$	&	$\langle p+\bar{p}\rangle_{acc}$	& $\langle p+\bar{p}\rangle$\\
	\hline
	31 	&	0.397(1)	&	3.556(37)	&	0.0440(3)	&	0.202(11)	&	0.280(1)	&	0.982(3)	\\
	40 	&	0.6233(3)	&	4.101(36)	&	0.0680(3)	&	0.254(11)	&	0.331(1)	&	1.101(3) \\
	80 	&	1.416(1)	&	4.701(38)	&	0.1563(3)	&	0.296(11)	&	0.369(1)	&	1.111(4)	\\
	158	&	2.360(2)	&	5.514(45)	&	0.2597(4)	&	0.366(18)	&	0.399(1)	&	1.194(10)\\
	\hline
	
\end{tabular}
\caption{Comparison of mean multiplicity in the analysis acceptance to mean multiplicity of identified hadrons in the full phase-space (only statistical uncertainty indicated)~\cite{Aduszkiewicz:2017sei}.}
\label{tab:means}
\end{table}

\begin{figure*}
\centering
\includegraphics[width=0.60\textwidth]{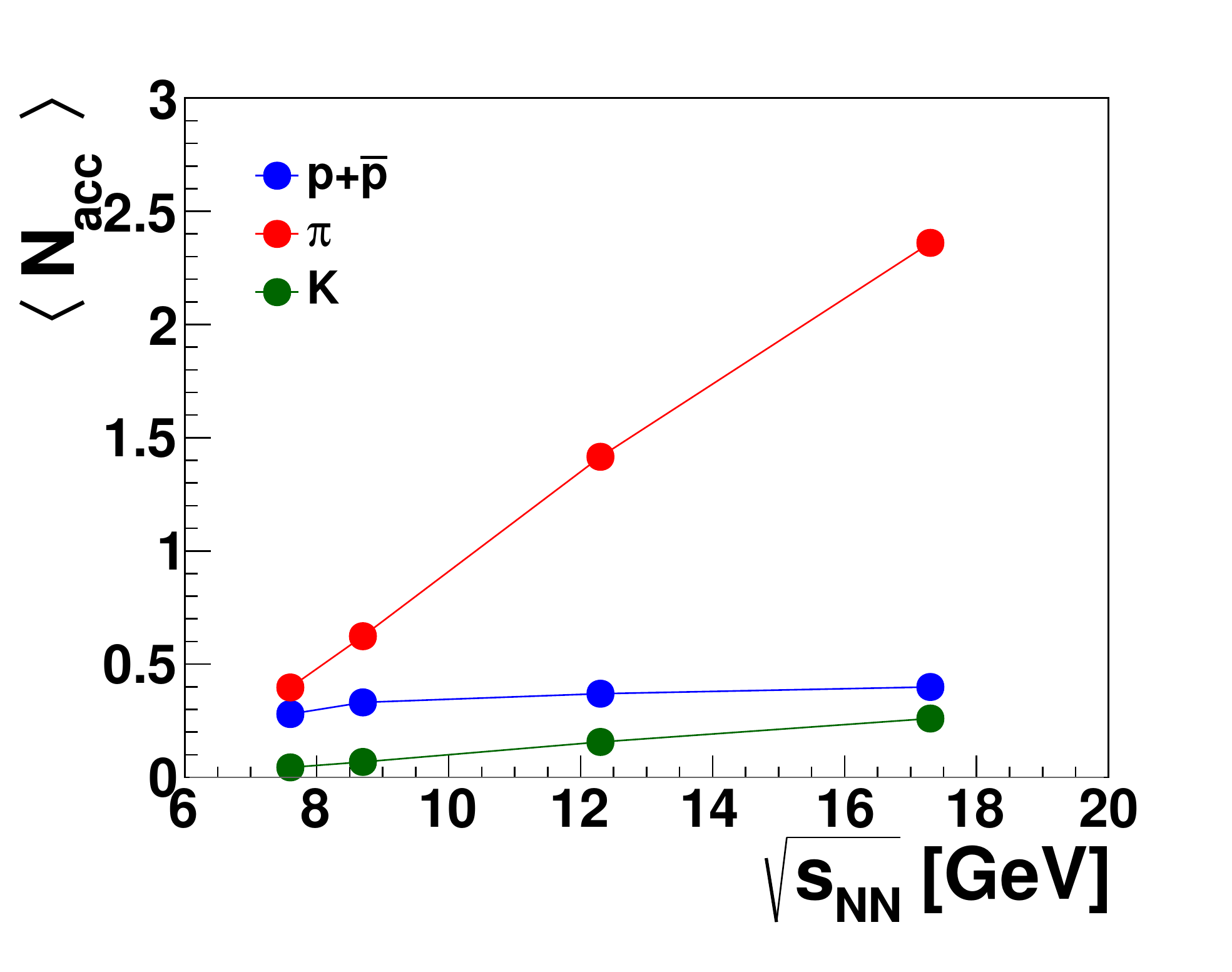}
\caption[]{
 Mean multiplicities of charged $\pi$, $K$ and $p+\bar{p}$
in the analysis acceptance as a function of collision energy.
Statistical uncertainties are smaller than the symbol size. Systematic uncertainties are not shown.} 
\label{fig:mult}
\end{figure*}

\begin{figure*}
	\centering
	\includegraphics[width=0.8\textwidth]{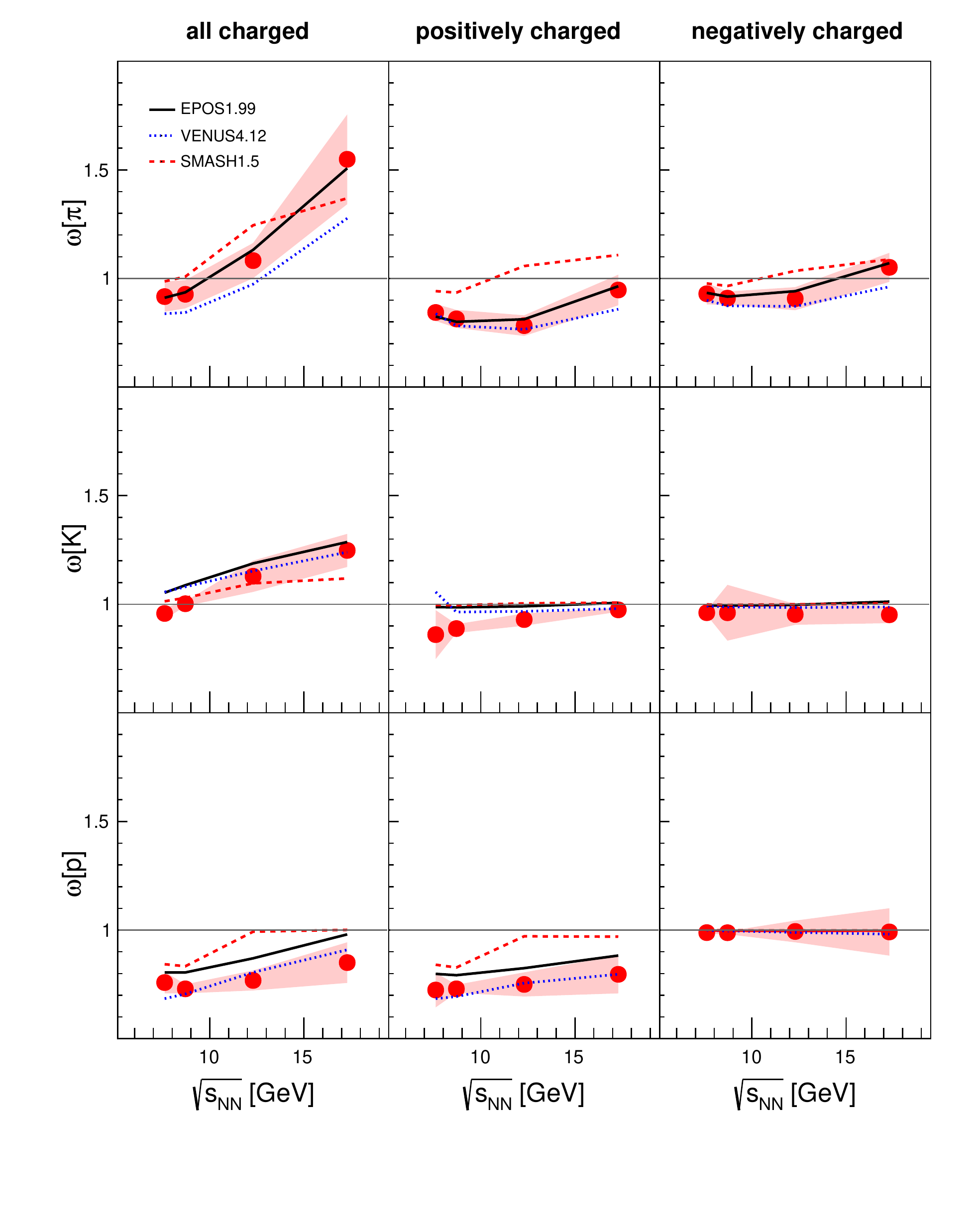}
				
	\caption[]{
	The collision energy dependence of scaled variance of pion, kaon
	and proton multiplicity distributions produced in inelastic p+p interactions. 
	Results for all charged, positively and negatively charged hadrons are presented separately.
	The solid, dashed and dotted lines show predictions of \Epos, \Smash and \Venus models, respectively.
    Statistical uncertainties are smaller than the symbol size
	and systematic uncertainty is indicated by red bands.
	}
\label{fig:omega}
\end{figure*}

\begin{figure*}
\centering
\includegraphics[width=0.8\textwidth]{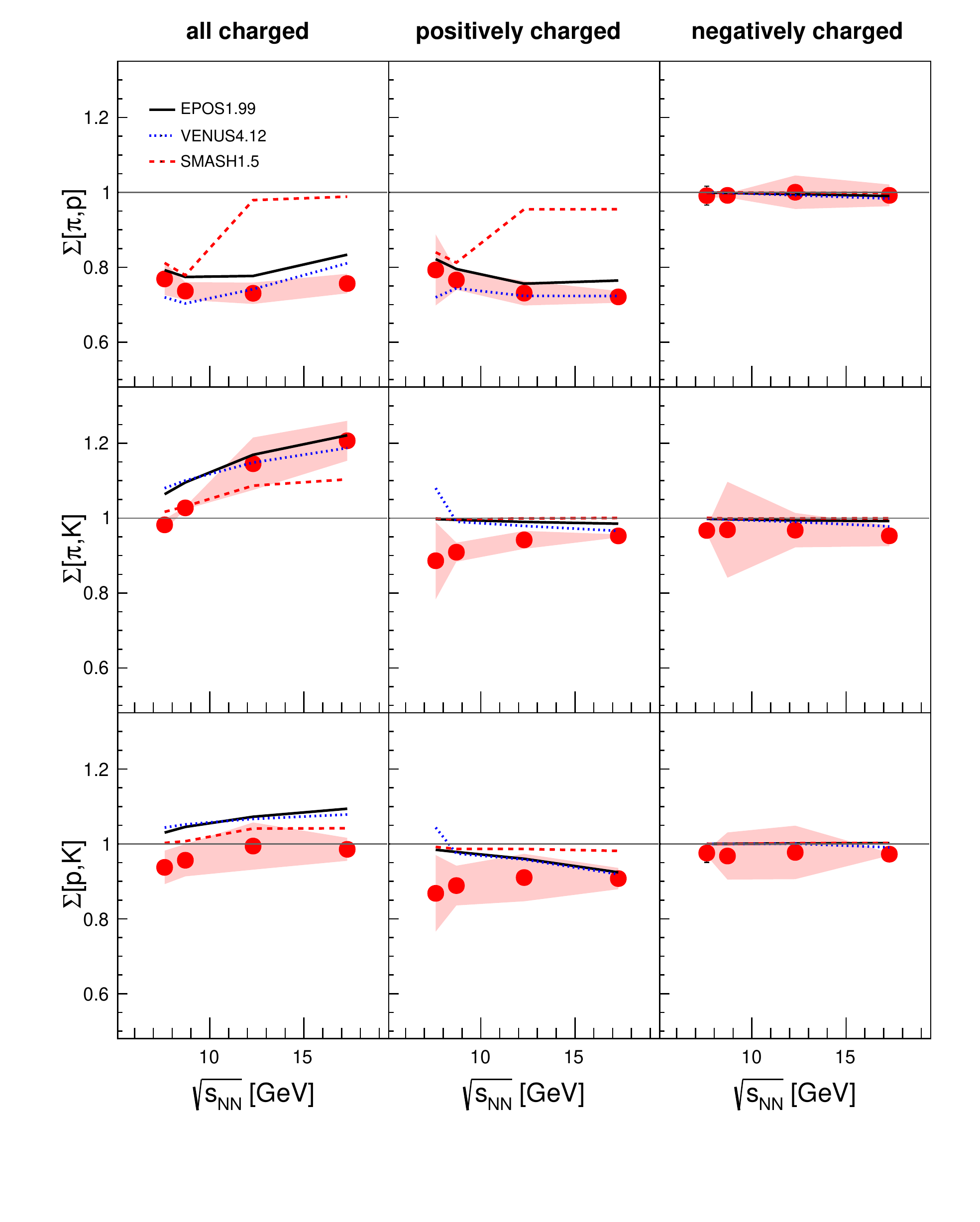}
				
\caption[]{
	The collision energy dependence of $\Sigma$ for different particle type combinations
	in inelastic p+p interactions. 
	Results for all charged, positively and negatively charged hadrons are presented separately.
		The solid, dashed and dotted lines show predictions of \Epos, \Smash and \Venus models, respectively. 
		Statistical uncertainties are smaller than the symbol size
		and systematic uncertainty is indicated by red bands.
	}
\label{fig:sigma}
\end{figure*}

\begin{figure*}
\centering
\includegraphics[width=0.8\textwidth]{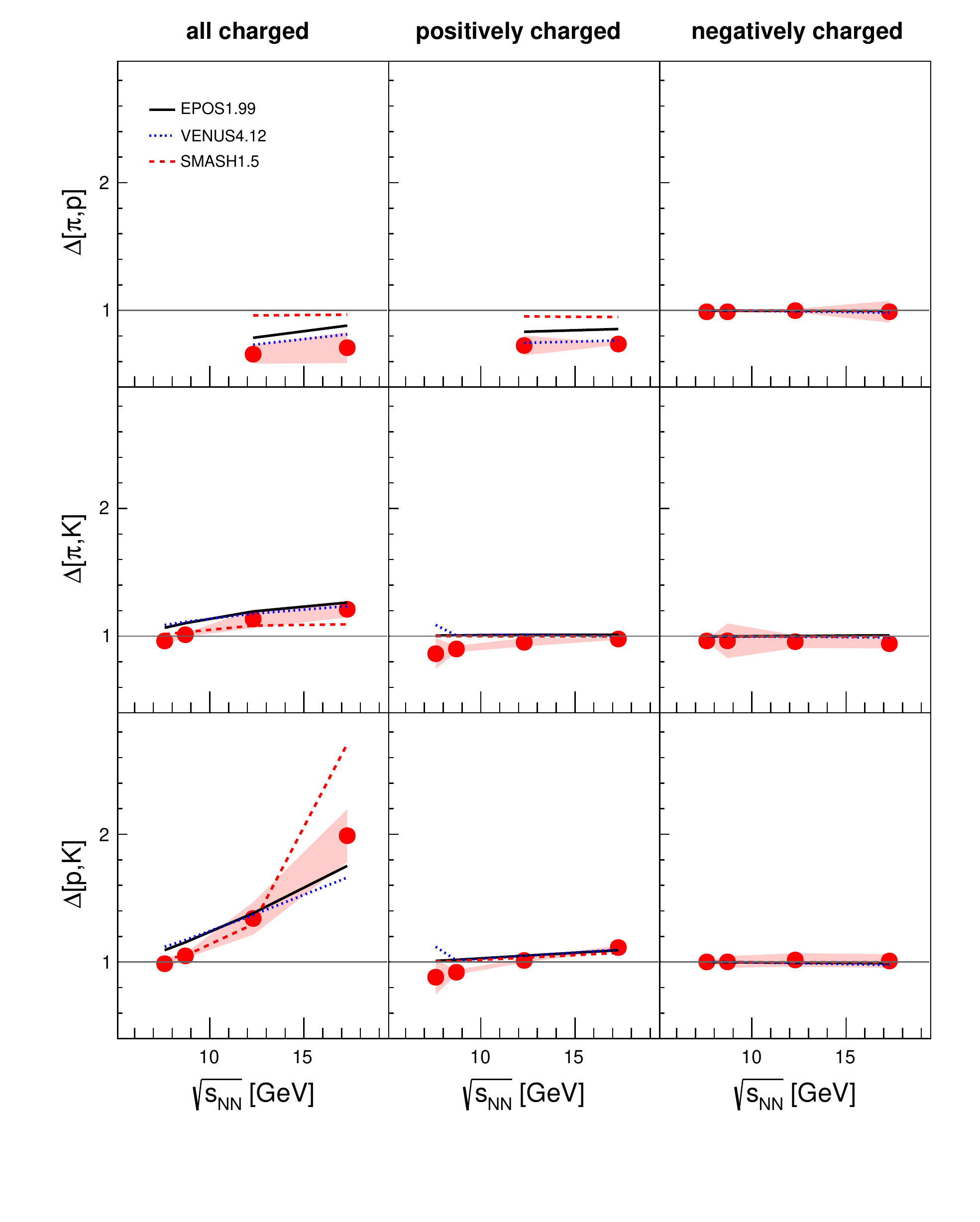}

\caption[]{
	The collision energy dependence of $\Delta$ for different particle type combinations
	in inelastic p+p interactions. 
	Results for all charged, positively and negatively charged hadrons are presented separately.
		The solid, dashed and dotted lines show predictions of \Epos, \Smash and \Venus models, respectively. 
		Statistical uncertainties are smaller than the symbol size
		 and systematic uncertainty is indicated by red bands.
}
\label{fig:delta}
\end{figure*}

\subsection{Results}
\label{sec:results:results}

The final results presented in this section refer to identified hadrons 
produced in inelastic p+p interactions by strong interaction processes and in  electromagnetic
decays of produced hadrons. They were obtained within the kinematic acceptances 
given in~Ref.~\cite{na61ParticlePopulationMatrix} and illustrated in Fig.~\ref{fig:acc-ptflu}. Note, that the kinematic acceptances for positively and negatively charged hadrons 
are different.

Mean multiplicities of pions, kaons and anti-protons in the acceptance region of the
fluctuation analysis are plotted in Fig.~\ref{fig:mult} and compared to corresponding mean multiplicities measured in the full phase-space in Table~\ref{tab:means}.

Pions are the most abundantly produced particles and are the majority of accepted charged hadrons in all analyzed reactions.
With decreasing beam momentum the contribution of protons increases and  
the small contributions of kaons and protons decrease.
Almost all charged hadrons are pions, except that protons 
are the majority of positively charged hadrons at the lowest beam momentum, 31~\GeVc.
The changes of particle type composition with charge of selected hadrons and beam momentum
are related to different thresholds for production of pions, kaons and anti-protons.
The mean proton multiplicity in the models in full phase-space is about one (0.3-0.4 in the acceptance) and approximately independent of beam momentum. This is because final state protons are strongly correlated with two initial state protons via baryon number conservation.


Figure~\ref{fig:omega} shows the collision energy dependence of
the scaled variance of pion, kaon and proton multiplicity distributions.
Note, the intensive fluctuation measure $\omega$ is one for a Poisson distribution and
zero in the case of a constant multiplicity for all collisions.
The scaled variance quantifies the width of the multiplicity distribution relatively to the width of the Poisson distribution with the same mean multiplicity.
The results for all charged, positively charged and negatively charged
hadrons are presented separately. 
One observes:
\begin{enumerate}[(i)]
\item  
	the scaled variance for pions increases with the collision energy. The increase is the strongest for all charged pions. This is probably related to the well established KNO scaling of the charged hadron multiplicity distributions in inelastic p+p interactions with the scaled variance being proportional to mean multiplicity~\cite{Koba:1972ng,Golokhvastov:2001ei,Golokhvastov:2001pt}.
	Global and local (resonance decays) electric charge conservation correlates multiplicities of positively and negatively charged pions and thus the effect is the most pronounced for all charged hadrons.
\item 
	The dependence of $\omega$ on beam momentum for kaons is qualitatively similar to the one for pions but weaker. This is probably related to the significantly smaller mean multiplicity of kaons than pions. One notes that the scaled variance of the multiplicity distribution approaches one when the  mean multiplicity decreases to zero.
	The latter effect is likely responsible for $\omega[\km]$ and $\omega[\pbar]$ being close to one, as the mean multiplicity of \km and \pbar in the acceptance is below 0.1 and 0.03, respectively.
\item 
    The scaled variance of protons is about 0.8 and depends weakly on the beam momentum.
    The net-baryon (baryon - anti-baryon) multiplicity in full phase-space is exactly two. This is because the initial baryon number is two and baryon number is conserved. Thus the scaled variance of the net-baryon multiplicity distribution is zero. Anti-baryon production at the SPS energies is 
    small and thus the net-baryon multiplicity is close to the baryon multiplicity.
    The baryons are predominately protons and neutrons. Thus the proton fluctuations are expected to be mostly due to the fluctuation of the proton to neutron ratio and fluctuations caused by the limited phase space acceptance of protons. 
\end{enumerate}


Figures~\ref{fig:sigma} shows the results on $\Sigma$ for 
pion-proton, pion-kaon and proton-kaon multiplicities
measured separately for all charged, positively
charged, and negatively charged hadrons produced in inelastic p+p collisions at 31--158~\GeVc beam momentum.
The $\Sigma$ measure assumes the value one in the Independent Particle Production Model which
postulates that particle types are attributed to particles independently of each other.
This implies that $\Sigma$ unlike $\omega$ is insensitive to the details of the particle multiplicity distribution.
One observes:
\begin{enumerate}[(i)]
\item 
    For all and positively charged  pions-proton combinations $\Sigma$ is significantly below one
    (approximately 0.8) and depends weakly on the beam momentum.
    This is likely due to a large fraction of pion-proton pairs coming from decays of
    baryonic resonances~\cite{Begun:2014boa,Gorenstein:2015ria}. 
    Corresponding results for Pb+Pb collisions were reported in
    Refs.~\cite{Anticic:2011am, Alt:2008ab, Schuster:2009ak}.
\item 
	$\Sigma$ for all charged pion-kaon combinations increases significantly with the beam momentum and is about 1.2 at 158~\GeVc. The origin of this behaviour is unclear.
\item
    For the remaining cases $\Sigma$ is somewhat below or close to one suggesting a small
    contribution of hadrons from resonance decays.
\end{enumerate}


Figure~\ref{fig:delta} shows the results for  $\Delta$ of identified hadrons calculated separately for all charged, positively
charged, and negatively charged hadrons produced in inelastic p+p collisions at beam momenta from 31 to 158~\GeVc. 
Results for  $\Delta[\pi,p]$ at 31 and 40~\GeVc are not shown since they have large
statistical uncertainties. This is  because for these reactions $N[\pi]\approx N[p]$ (see Fig.~\ref{fig:mult}) and thus  $C_{\Delta}\approx 0$ (see Eq.~\ref{Eq:siq.delta}.)
The general properties of $\Delta$ are similar to the properties of $\Sigma$ discussed above.
Unlike $\Sigma$, $\Delta$ does not include a correlation term between multiplicities of 
two hadron types, see Eqs.~\ref{Eq:siq.delta} and~\ref{Eq:siq.sigma}.
One observes:
\begin{enumerate}[(i)]
\item
    $\Delta$ for all and positively charged pions and protons is below one.
    This is qualitatively similar to $\Sigma$ and thus likely to be caused by resonance decays.
\item 
	$\Delta[(p+\bar{p}),K]$ increases with the collision energy from about one to two.
	The origin of this dependence is unclear.
\end{enumerate}

\subsection{Comparison with models}

The results shown  in Figs.~\ref{fig:omega},~\ref{fig:sigma}
and~\ref{fig:delta}  are compared with predictions of three string-resonance models:
\Epos~\cite{Werner:2008zza,crmc} (solid lines), \SmashLong (dashed lines)~\cite{Weil:2016zrk} 
and \Venus~\cite{Werner:1993uh, Werner:1990aa} (dotted lines).

These models define the baseline for heavy ion collisions from which 
any critical phenomena are expected to emerge. However, the models should first be tuned to 
the experimental data on p+p interactions presented here.
In p+p interactions at CERN SPS energies one expects none of the high matter density phenomena usually
studied and searched for in nucleus-nucleus collisions. Any deviations from independent particle production
are considered to be caused by well established effects discussed in Sec.~\ref{sec:results:results}.
In general, the \Epos and \Venus models reproduce the results reasonably well. 
However, none of the models agrees with all features of the presented results. For a number a number of observables qualitative disagreement is observed for \Smash and \Venus.

	\section{Summary and outlook}

In this paper experimental results on multiplicity fluctuations of
identified hadrons produced in inelastic p+p interactions at 31, 40, 80,
and 158~\GeVc beam momentum are presented.  
Results were corrected for incomplete particle identification using 
a data-based procedure - the Identity method.
Remaining biases were corrected for utilising full physics and detector response simulations.
The sub-sample method was used to calculate statistical uncertainties whereas systematic
uncertainties were estimated by changing event and track selection cuts as well as
models.

Results on the scaled variance of multiplicity fluctuations $\omega[a]$ of pions, kaons and protons 
for all charged, positively charged and negatively charged hadrons are presented. 
Moreover results on the strongly intensive measures of multiplicities fluctuations of
two hadron types $\Sigma[a,b]$ and $\Delta[a,b]$ are shown.
These were obtained for pion-proton, pion-kaon and proton-kaon particle type combinations for
all charged as well as separately for positively and negatively charged pair combinations.
The results are presented as a function of the collision energy and discussed in the
context of KNO-scaling, conservation laws and resonance decays. 

Finally, the \NASixtyOne measurements are compared with string-resonance models \Smash, \Epos and \Venus.
In general, the \Epos and \Venus models reproduce the results reasonably well. 
However, none of the models agree with all presented results. For some observables even 
qualitative disagreement is observed for the \Smash and \Venus models. 
Thus, before the models can provide the baseline for heavy ion collisions 
in the search for critical phenomena, the models need to be tuned to the experimental data on p+p interactions
presented in this paper.

	\appendix
\section{Details on the Identity Method}
\label{sec:app2}
\label{sec:smeard}

\begin{figure*}
	\centering

	\includegraphics[width=0.3\textwidth]{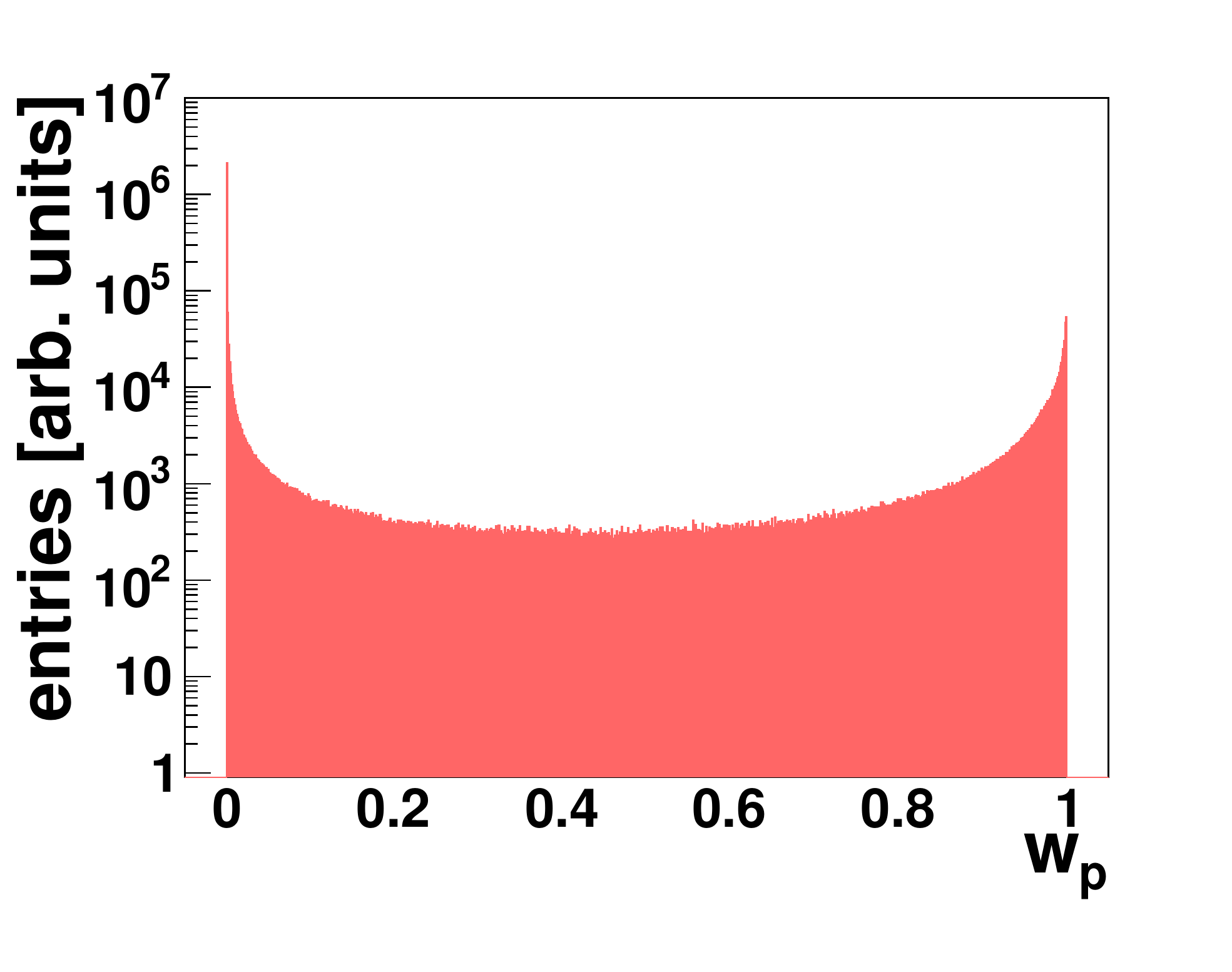}
	\quad
	\includegraphics[width=0.3\textwidth]{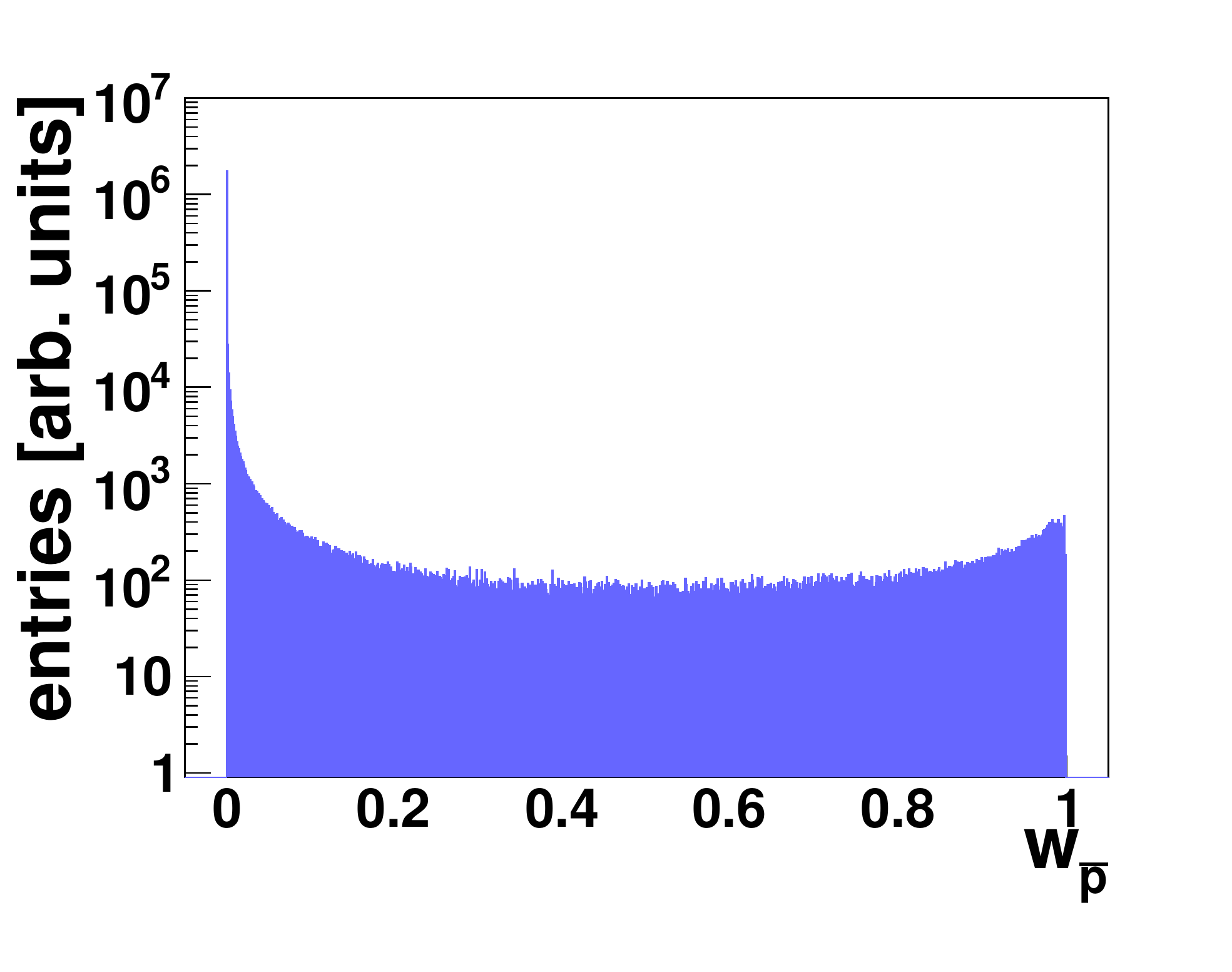}
		\\
	\includegraphics[width=0.3\textwidth]{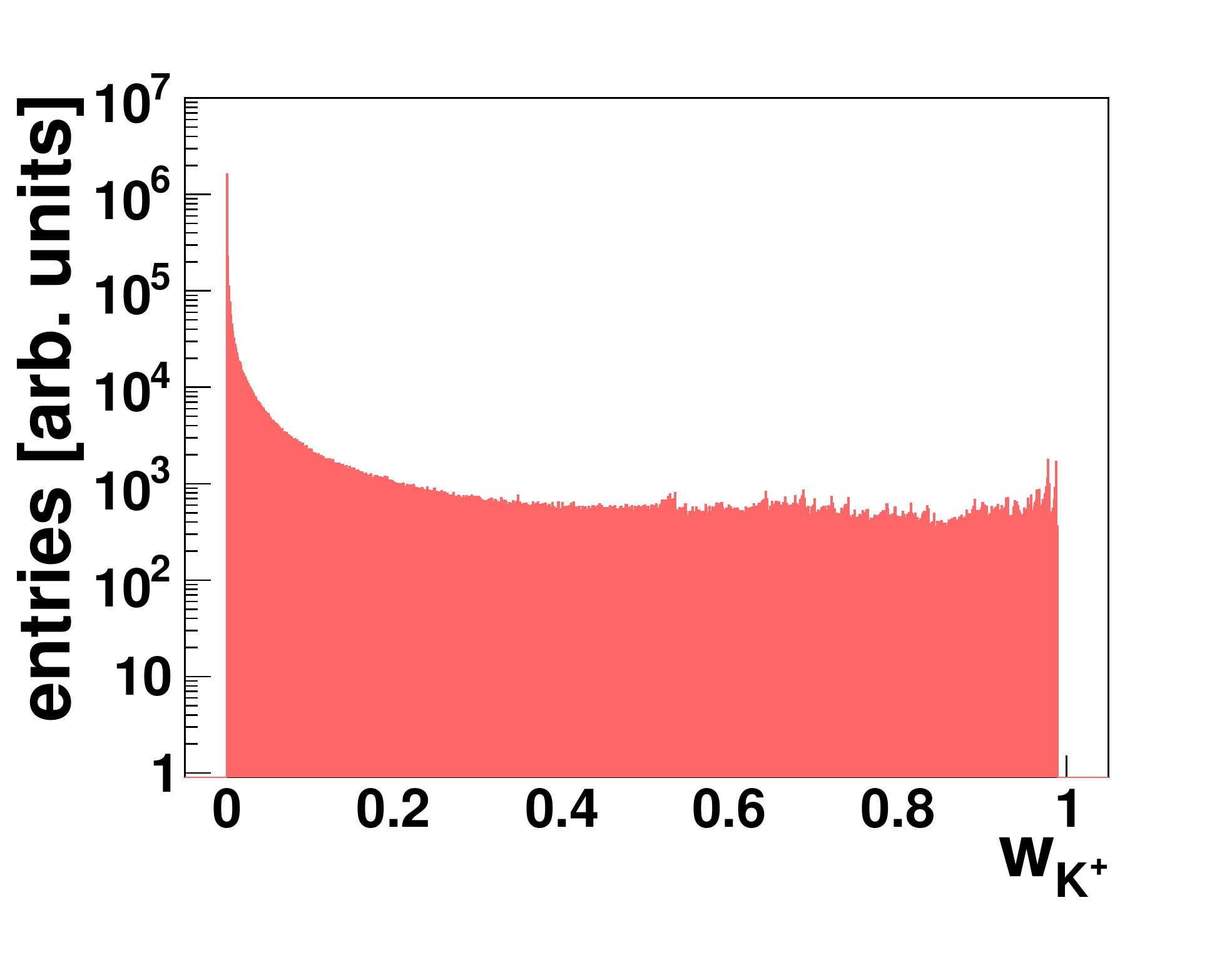}
	\quad
	\includegraphics[width=0.3\textwidth]{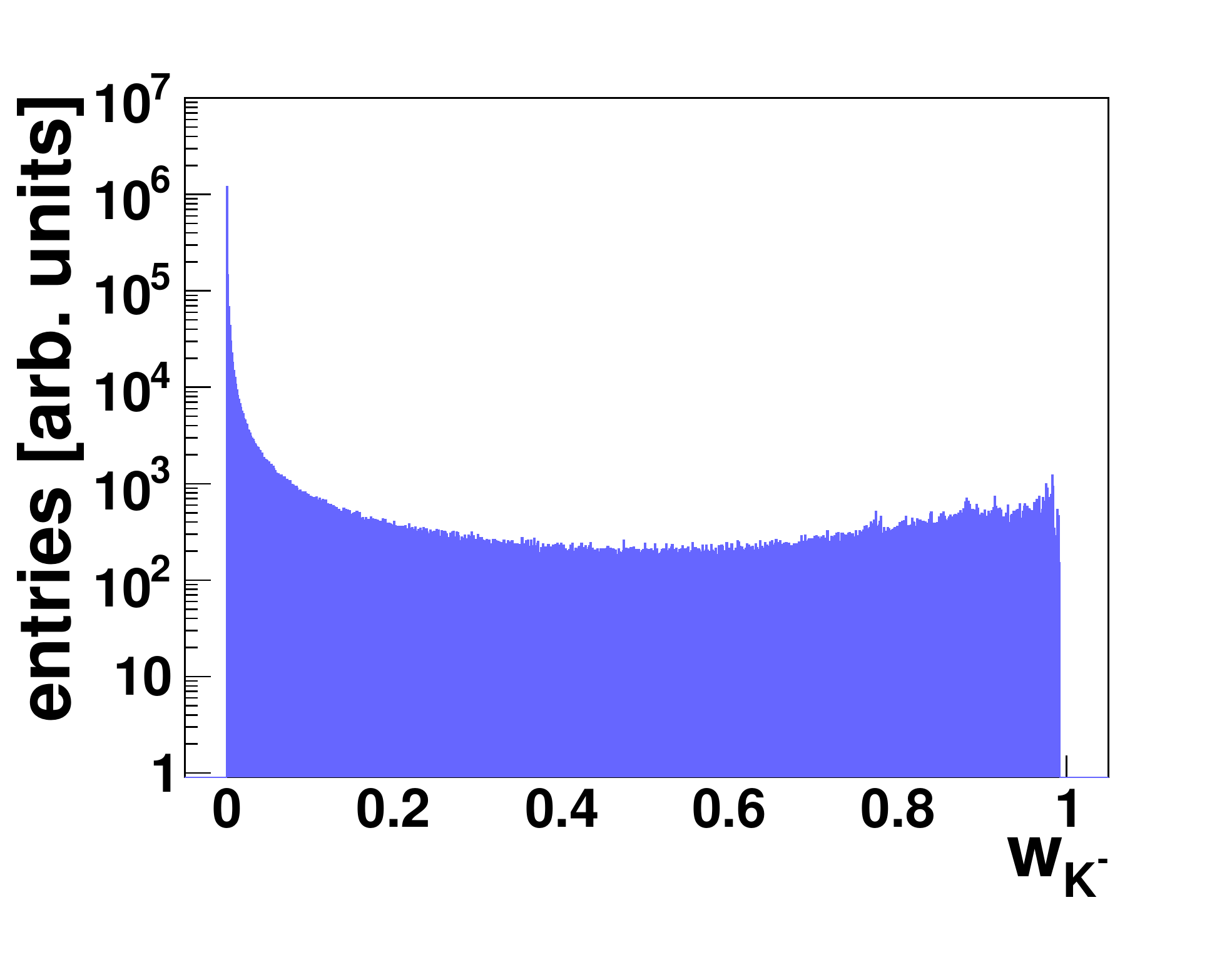}
		\\
	\includegraphics[width=0.3\textwidth]{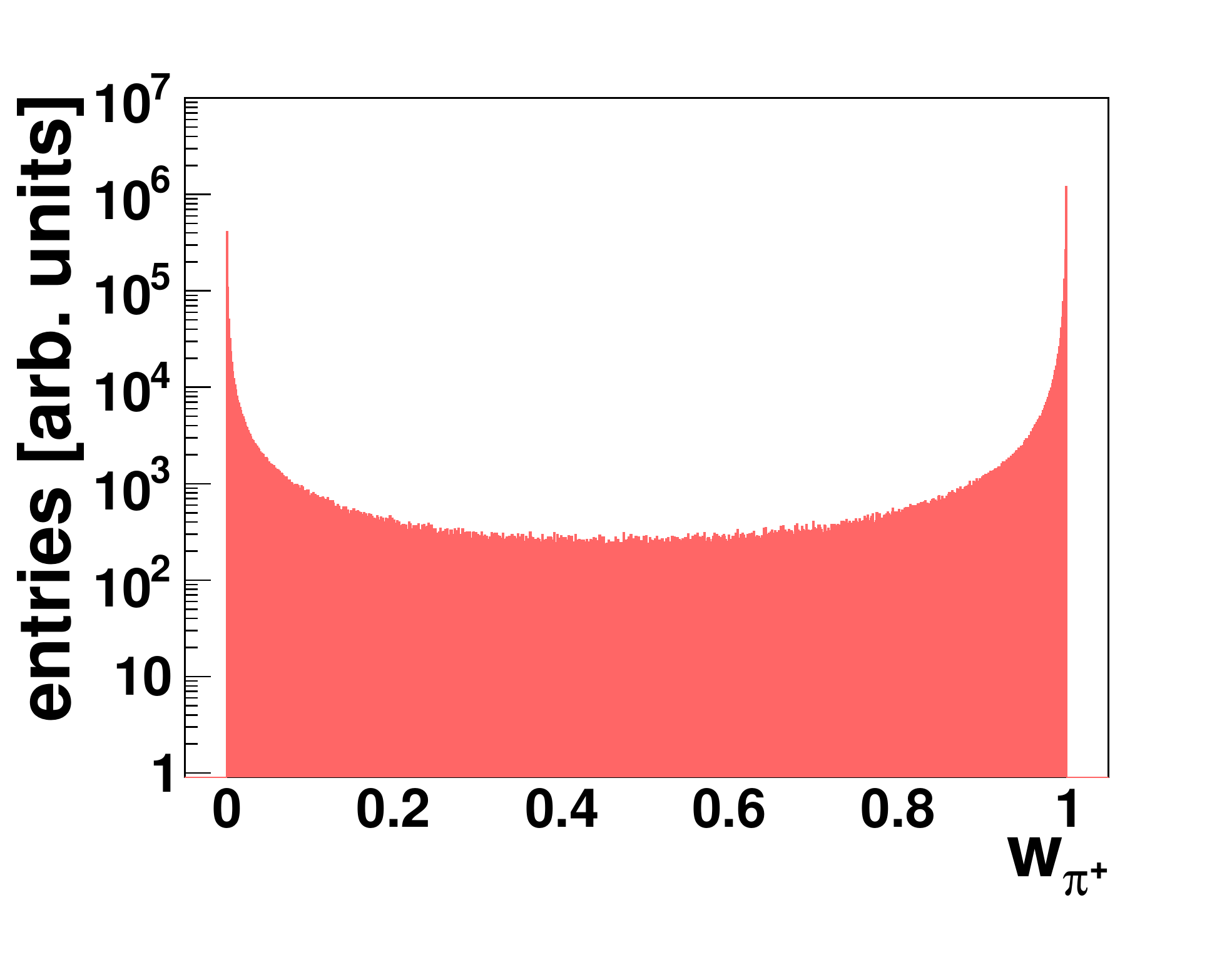}
	\quad
	\includegraphics[width=0.3\textwidth]{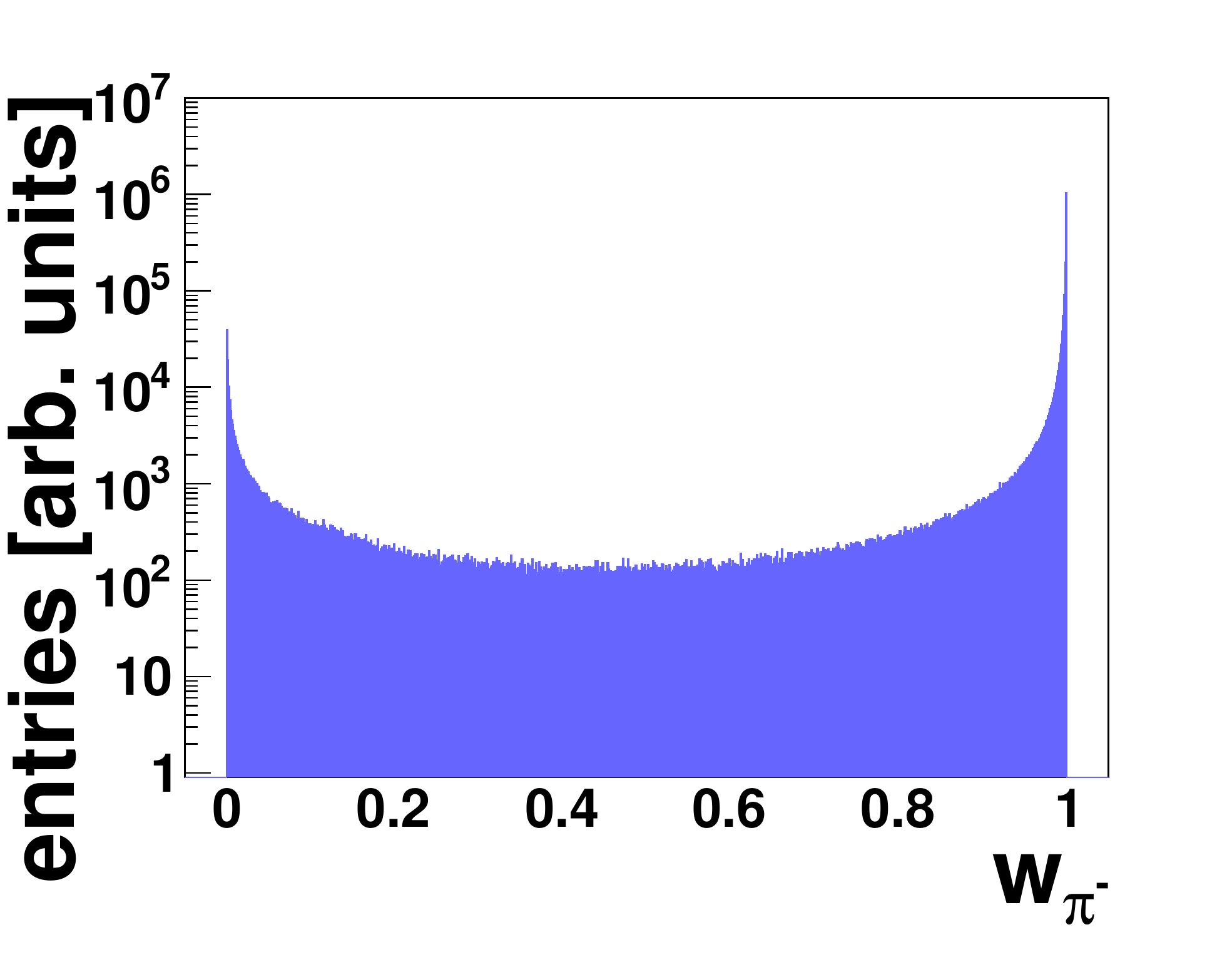}
	\\
	\includegraphics[width=0.3\textwidth]{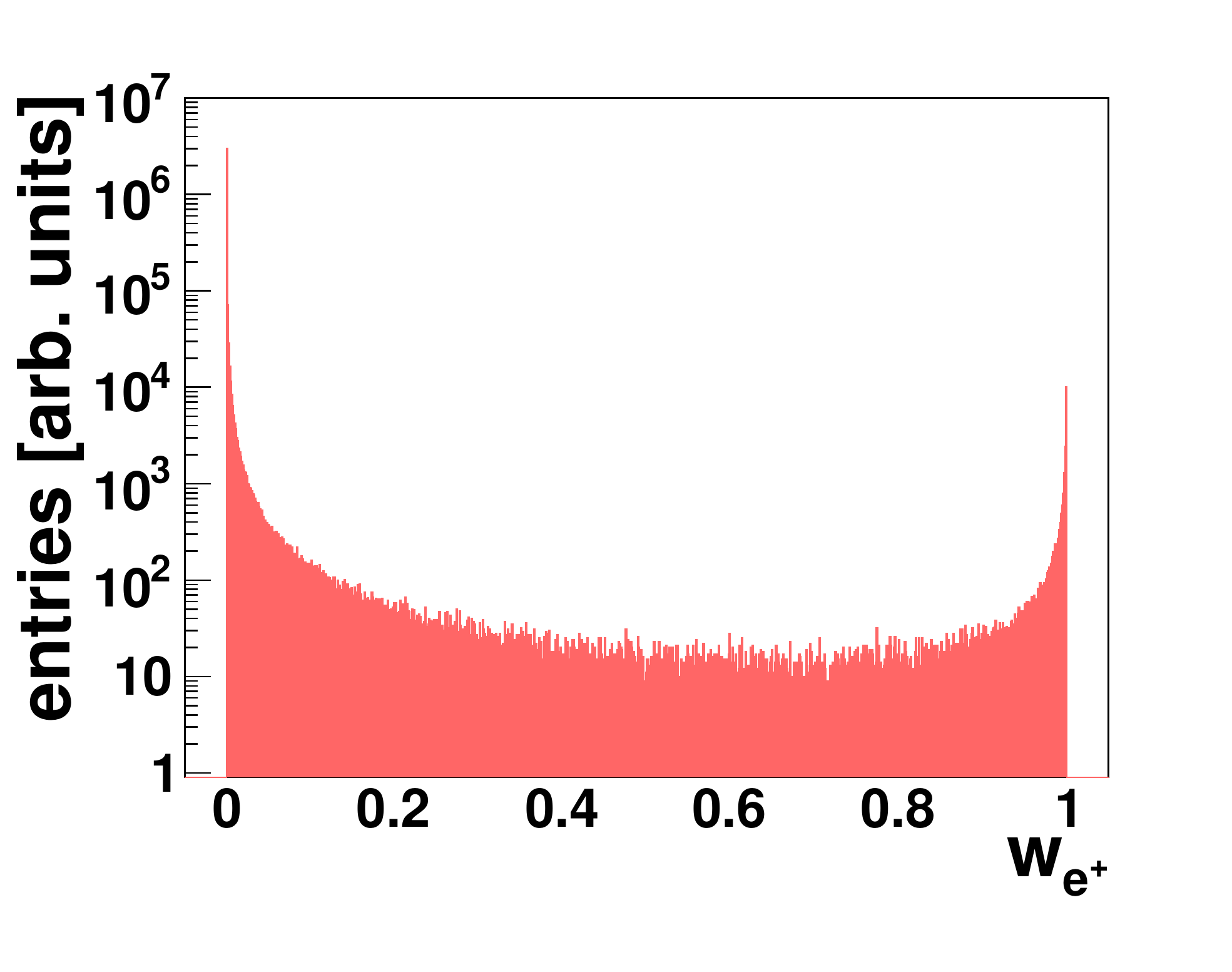}
	\quad
	\includegraphics[width=0.3\textwidth]{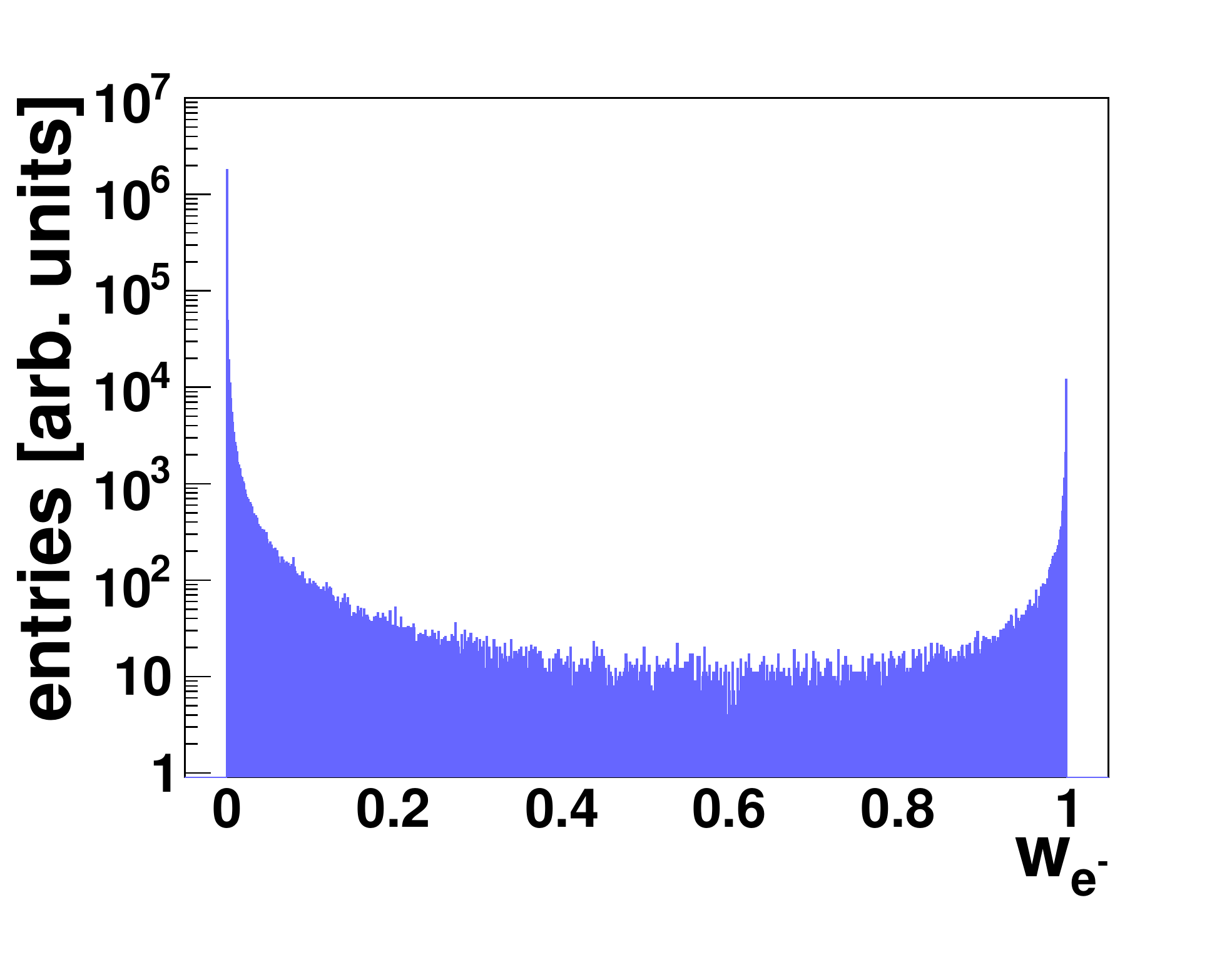}

	\caption[]{
			Distributions of probabilities of positively (\textit{left}) and negatively
			(\textit{right}) charged
			particles (from top to bottom: \textit{p}, \textit{K}, $\pi$, \textit{e}) selected for the analysis in p+p interactions at
			158~\GeVc.
	}
	\label{fig:identities158}
	\end{figure*}

The parametrization of inclusive \dedx spectra of identified particles is first used to calculate
probabilities w$_a$ (see Sec.~\ref{sec:identity}).
Distributions of w$_a$ for p+p interactions at 
158~\GeVc are shown in Fig.~
\ref{fig:identities158} for positively and
negatively charged particles, separately.

\begin{figure}
	\centering
		\includegraphics[width=0.3\textwidth]{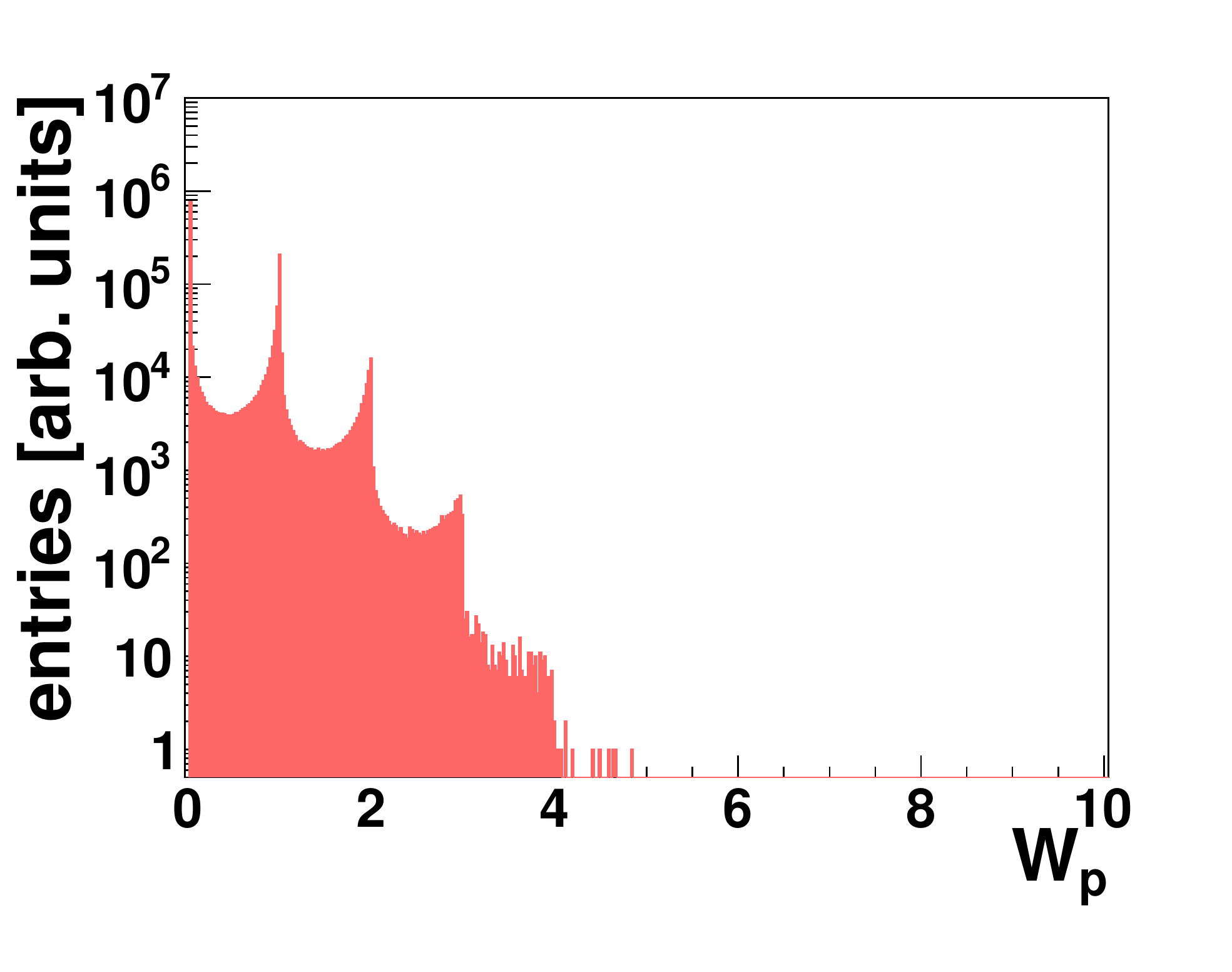}
	\quad
	\includegraphics[width=0.3\textwidth]{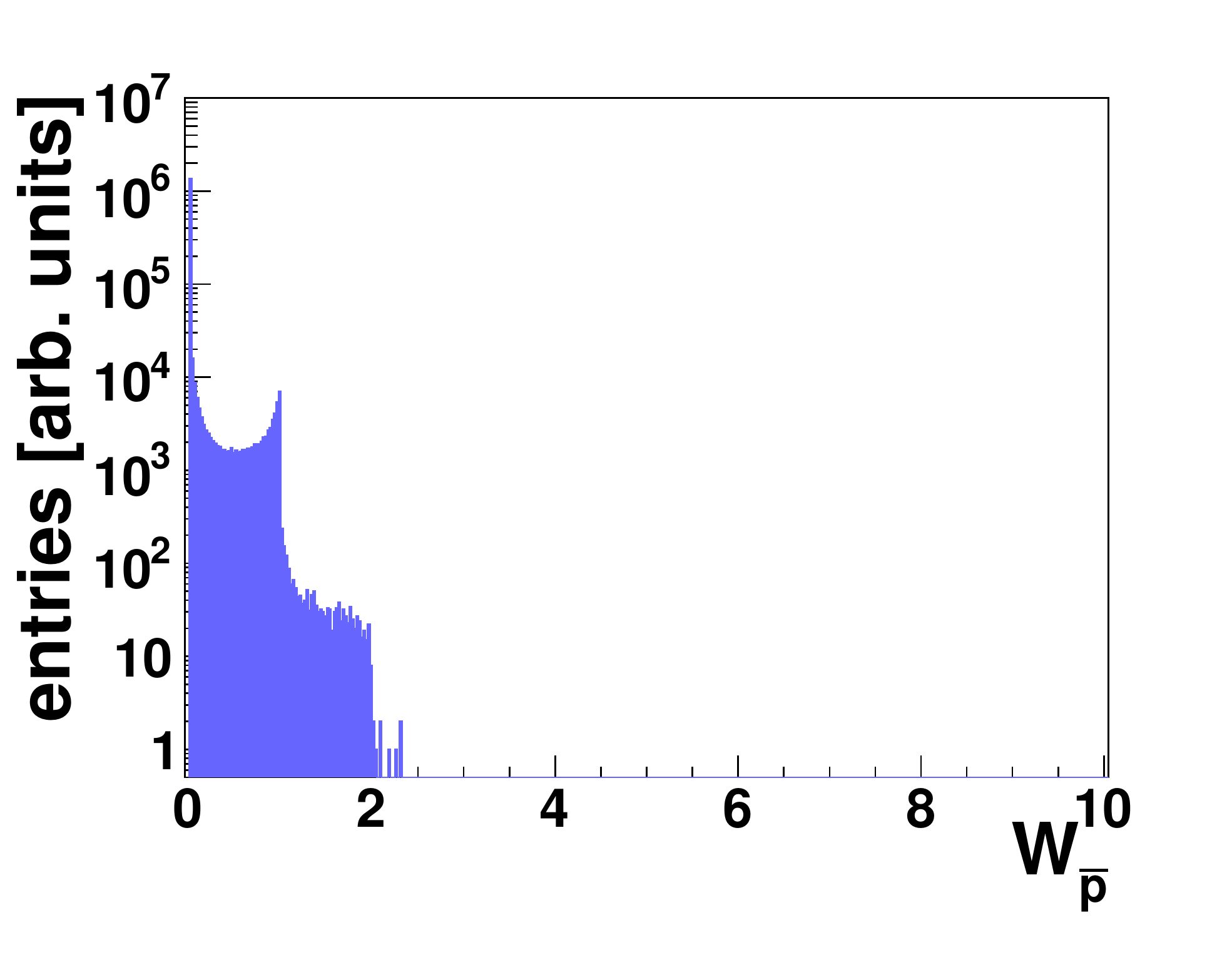}
		\\
	\includegraphics[width=0.3\textwidth]{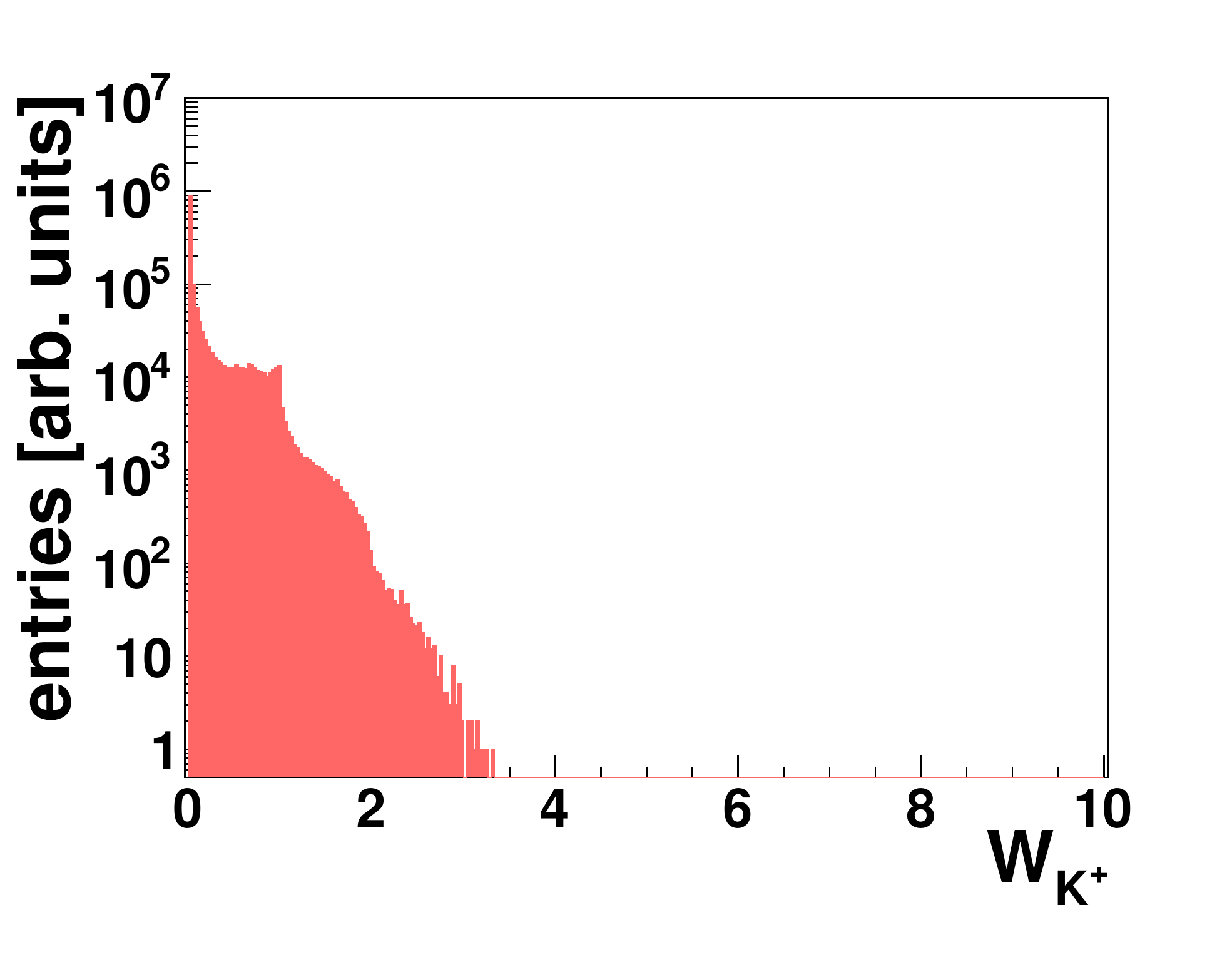}
	\quad
	\includegraphics[width=0.3\textwidth]{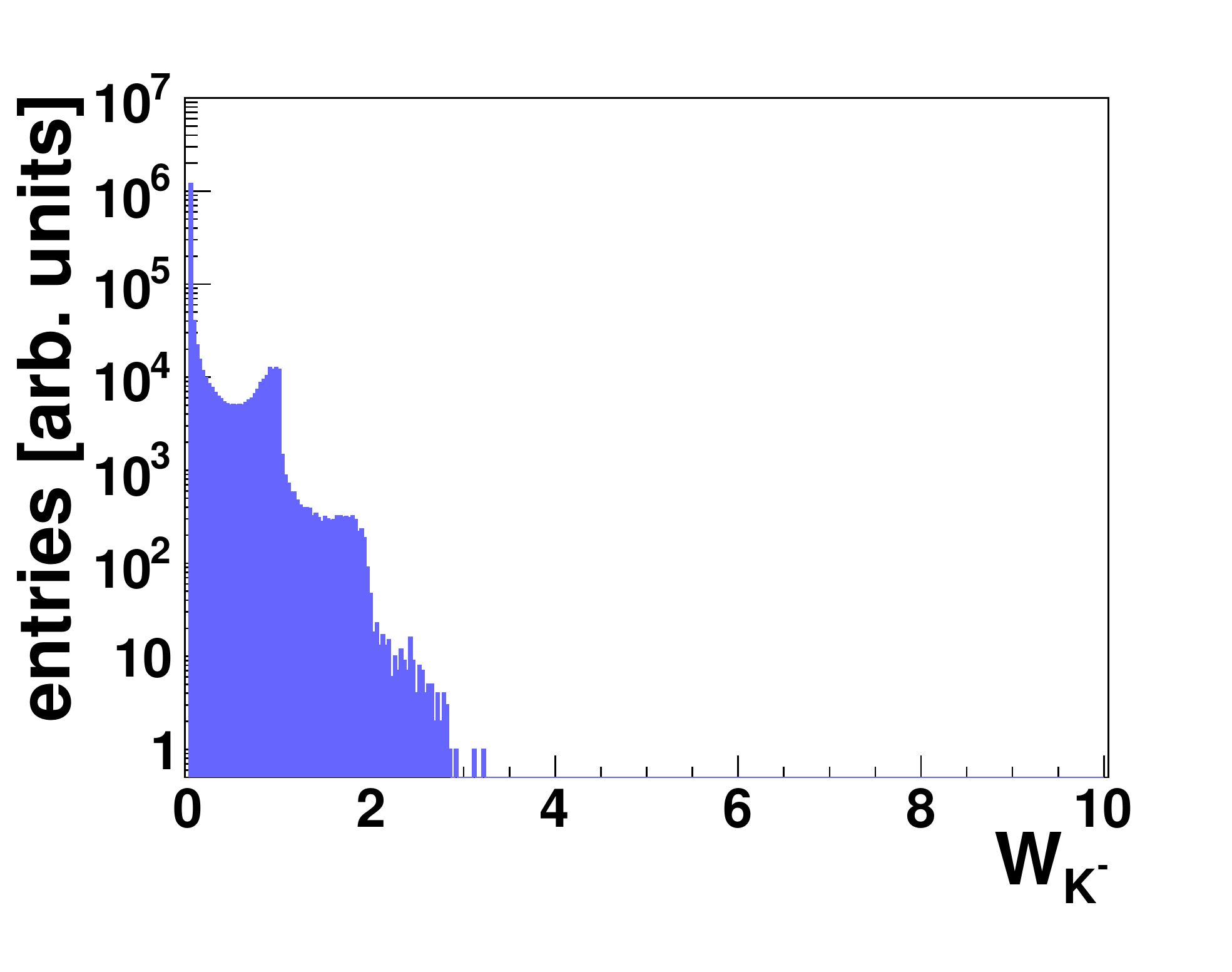}
		\\
	\includegraphics[width=0.3\textwidth]{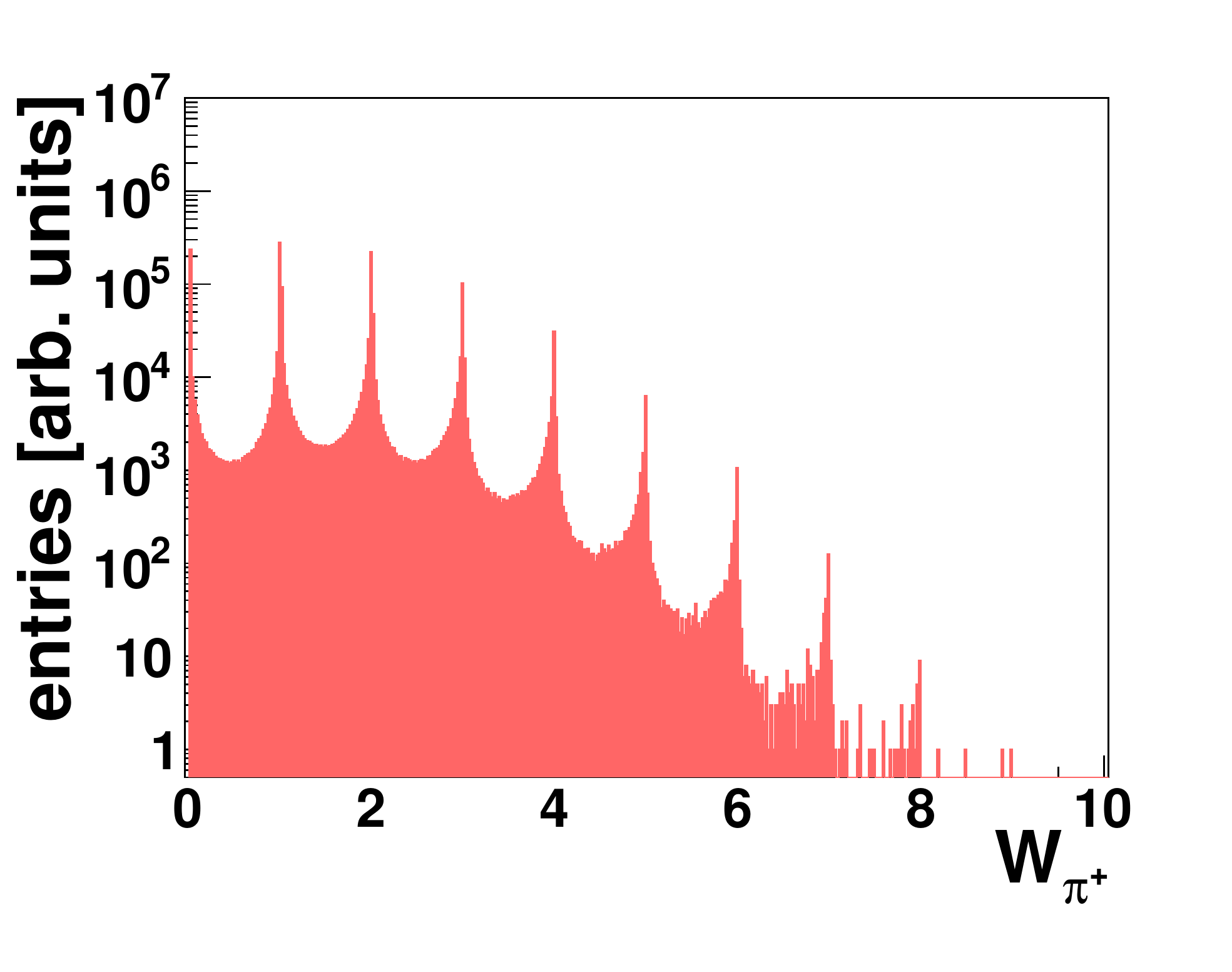}
	\quad
	\includegraphics[width=0.3\textwidth]{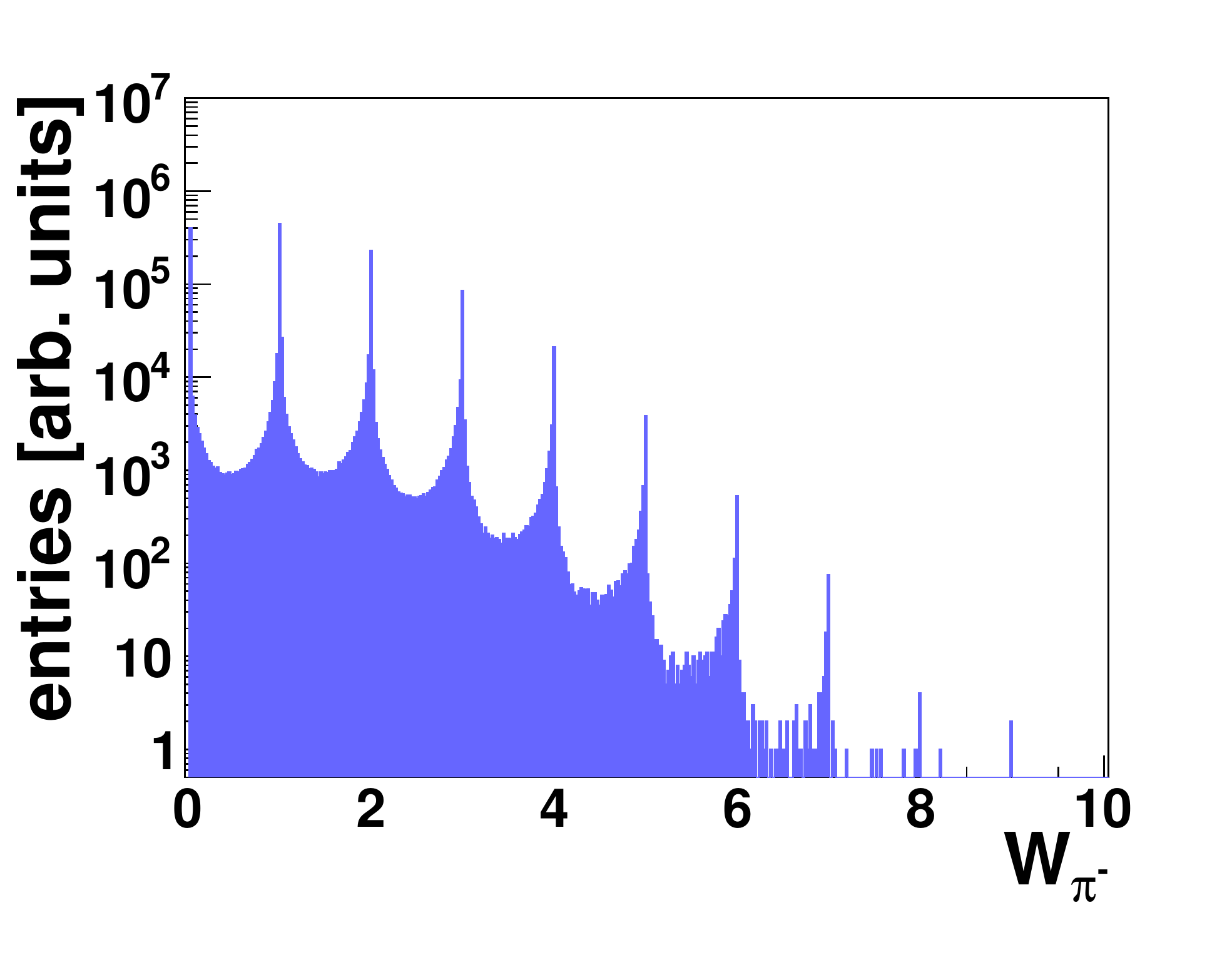}
	\\
	\includegraphics[width=0.3\textwidth]{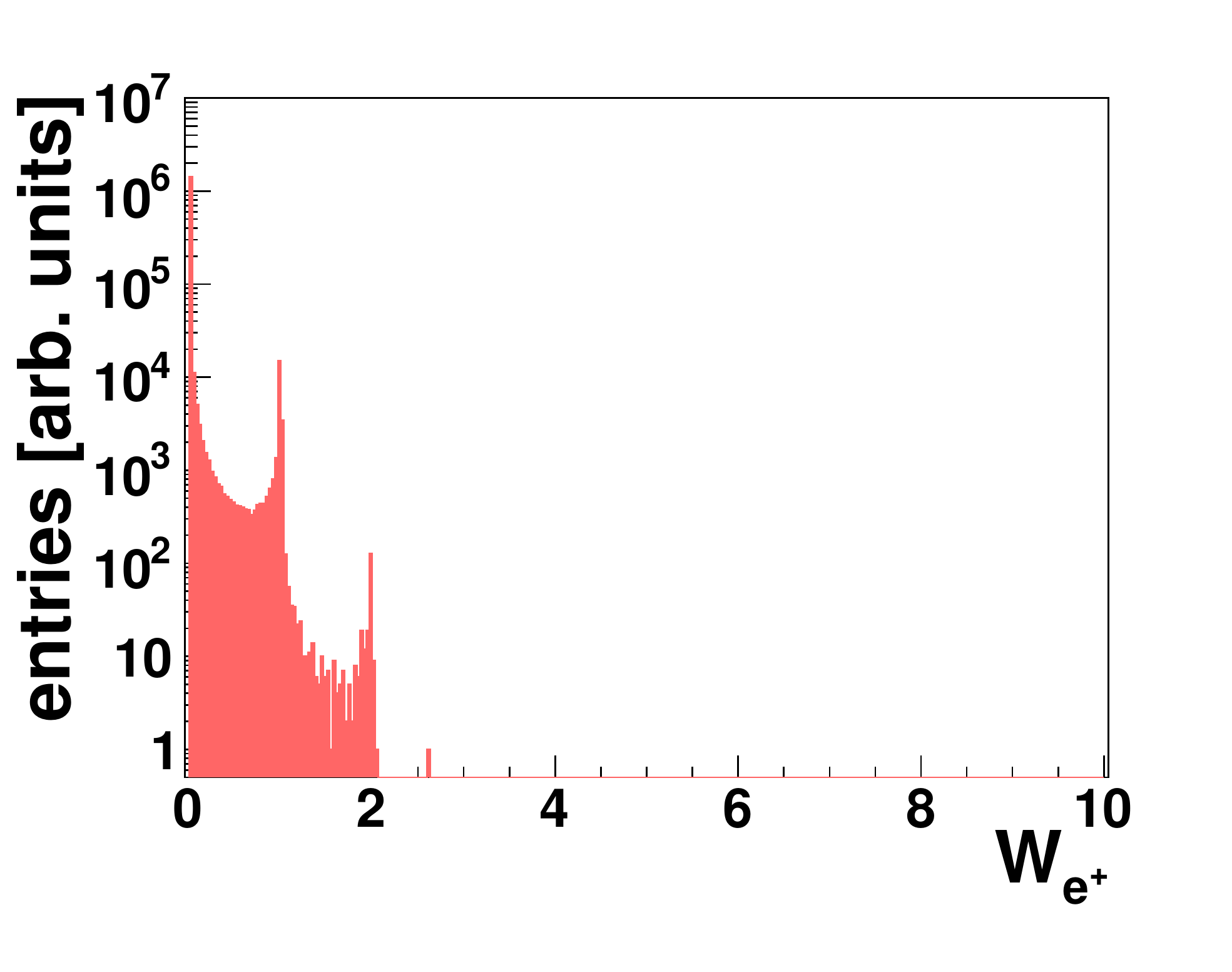}
	\quad
	\includegraphics[width=0.3\textwidth]{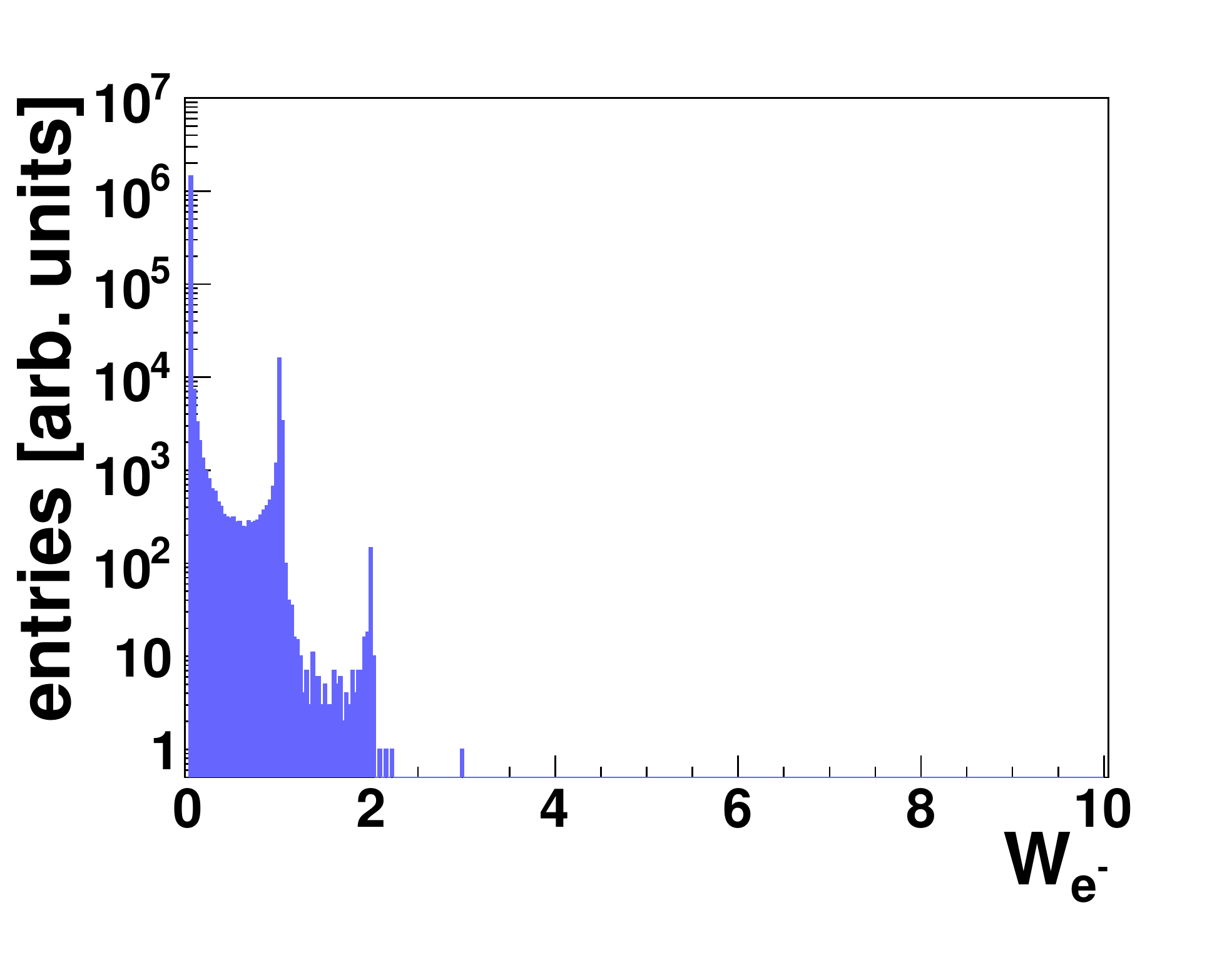}

	\caption[]{
		Smeared multiplicity
		distributions of positively (\textit{left}) and negatively
		(\textit{right}) charged particles (from \textit{top} to \textit{bottom}: \textit{p}, \textit{K}, $\pi$, \textit{e})
		in p+p interactions at 158~\GeVc.
	}
	\label{fig:W158}
\end{figure}

In the second step smeared multiplicities of identified particles W$_a$
(see Sec.~\ref{sec:identity}) are calculated for each selected event and their distributions are obtained.
Examples of smeared multiplicity distributions for p+p interactions at 
158~\GeVc are shown
in Fig.~
\ref{fig:W158} for positively and negatively charged particles, separately.

Finally, first and second moments of smeared multiplicity distributions are calculated
for positively and negatively charged particles, separately.

\clearpage
\section*{Acknowledgements}
We would like to thank the CERN EP, BE, HSE and EN Departments for the
strong support of NA61/SHINE.

This work was supported by
the Hungarian Scientific Research Fund (grant NKFIH 123842\slash123959),
the Polish Ministry of Science
and Higher Education (grants 667\slash N-CERN\slash2010\slash0,
DIR\slash WK\slash\- 2016\slash 2017\slash\- 10-1,
NN\,202\,48\,4339 and NN\,202\,23\,1837), the National Science Centre Poland(grants
2014\slash14\slash E\slash ST2\slash00018,
2014\slash15\slash B\slash ST2 \slash\- 02537 and
2015\slash18\slash M\slash ST2\slash00125,
2015\slash 19\slash N\slash ST2 \slash01689,
2016\slash23\slash B\slash ST2\slash00692,
2017\slash\- 25\slash N\slash\- ST2\slash\- 02575,
2018\slash 30\slash A\slash ST2\slash 00226,
2018\slash 31\slash G\slash ST2\slash 03910),
WUT ID-UB,
the Russian Science Foundation, grant 16-12-10176 and 17-72-20045,
the Russian Academy of Science and the
Russian Foundation for Basic Research (grants 08-02-00018, 09-02-00664
and 12-02-91503-CERN),
the Russian Foundation for Basic Research (RFBR) funding within the research project no. 18-02-40086,
the National Research Nuclear University MEPhI in the framework of the Russian Academic Excellence Project (contract No.\ 02.a03.21.0005, 27.08.2013),
the Ministry of Science and Higher Education of the Russian Federation, Project "Fundamental properties of elementary particles and cosmology" No 0723-2020-0041,
the European Union's Horizon 2020 research and innovation programme under grant agreement No. 871072,
the Ministry of Education, Culture, Sports,
Science and Tech\-no\-lo\-gy, Japan, Grant-in-Aid for Sci\-en\-ti\-fic
Research (grants 18071005, 19034011, 19740162, 20740160 and 20039012),
the German Research Foundation (grant GA\,1480/8-1), the
Bulgarian Nuclear Regulatory Agency and the Joint Institute for
Nuclear Research, Dubna (bilateral contract No. 4799-1-18\slash 20),
Bulgarian National Science Fund (grant DN08/11), Ministry of Education
and Science of the Republic of Serbia (grant OI171002), Swiss
Nationalfonds Foundation (grant 200020\-117913/1), ETH Research Grant
TH-01\,07-3 and the Fermi National Accelerator Laboratory (Fermilab), a U.S. Department of Energy, Office of Science, HEP User Facility managed by Fermi Research Alliance, LLC (FRA), acting under Contract No. DE-AC02-07CH11359 and the IN2P3-CNRS (France).

\bibliographystyle{include/na61Utphys}


\bibliography{include/na61References}

\newpage
{\Large The \NASixtyOne Collaboration}
\bigskip
\begin{sloppypar}

\noindent
A.~Acharya$^{\,9}$,
H.~Adhikary$^{\,9}$,
A.~Aduszkiewicz$^{\,15}$,
K.K.~Allison$^{\,25}$,
E.V.~Andronov$^{\,21}$,
T.~Anti\'ci\'c$^{\,3}$,
V.~Babkin$^{\,19}$,
M.~Baszczyk$^{\,13}$,
S.~Bhosale$^{\,10}$,
A.~Blondel$^{\,4}$
% ...

\end{sloppypar}

\noindent
% ...
$^{3}$~Ru{\dj}er Bo\v{s}kovi\'c Institute, Zagreb, Croatia\\
$^{4}$~LPNHE, University of Paris VI and VII, Paris, France\\
% ...
$^{9}$~Jan Kochanowski University in Kielce, Poland\\
$^{10}$~Institute of Nuclear Physics, Polish Academy of Sciences, Cracow, Poland\\
% ...
$^{13}$~AGH - University of Science and Technology, Cracow, Poland\\
% ...
$^{15}$~University of Warsaw, Warsaw, Poland\\
% ...
$^{19}$~Joint Institute for Nuclear Research, Dubna, Russia\\
% ...
$^{21}$~St. Petersburg State University, St. Petersburg, Russia\\
% ...
$^{25}$~University of Colorado, Boulder, USA\\
% ...

\end{document}